\documentclass[a4paper,12pt]{article}
\usepackage[a4paper,text={16.8cm,22.4cm}]{geometry}
\usepackage{amsmath,amssymb,bm,braket,psfrag,colortbl,booktabs,latexsym}
\usepackage{graphicx}
\usepackage{color}
\usepackage{cite}
\usepackage{fancyhdr}

\allowdisplaybreaks 
\newcommand{\Tr}{\mathrm{Tr}}

\newcommand{\ff}{f\hspace{-0.4em}f}

\newcommand{\tNNLL}{\mbox{{\tiny NNLL}}}
\newcommand{\tNLL}{\mbox{{\tiny NLL}}}
\newcommand{\tNLO}{\mbox{{\tiny NLO}}}

\newcommand{\be}{\begin{equation}}
\newcommand{\ee}{\end{equation}}
\newcommand{\st}{\tilde{t}_1}

\begin{document}

\begin{titlepage}

  \begin{flushright}
    {MITP/13-85}\\
    {PSI-PR-13-15}\\
    December 16, 2013
  \end{flushright}
  
  \vspace{5ex}
  
  \begin{center}
    \textbf{\Large NNLL Momentum-Space Resummation for\\
            Stop-Pair Production at the LHC} \vspace{7ex}
    
    \textsc{Alessandro Broggio$^b$, Andrea Ferroglia$^c$, Matthias Neubert$^{d,e}$,\\
     Leonardo Vernazza$^f$ and Li Lin Yang$^{a,g,h}$} \vspace{2ex}
  
    \textsl{
      ${}^a$School of Physics and State Key Laboratory of Nuclear Physics and Technology\\
            Peking University, 100871 Beijing, China\\[0.3cm]
      ${}^b$Paul Scherrer Institute\\ 
            CH-5232 Villigen, Switzerland\\[0.3cm]
      ${}^c$New York City College of Technology, The City University of New York,\\ 
            300 Jay Street, Brooklyn, NY 11201, U.S.A.\\[0.3cm]
      ${}^d$PRISMA Cluster of Excellence \& Mainz Institut for Theoretical Physics\\
            Johannes Gutenberg University, D-55099 Mainz, Germany\\[0.3cm]
      ${}^e$Department of Physics, LEPP, Cornell University, Ithaca, NY 14853, U.S.A.\\[0.3cm]
      ${}^f$Dipartimento di Fisica, Universit\`a di Torino \& INFN, Sezione di Torino\\
            Via P.\ Giuria 1, I-10125 Torino, Italy\\[0.3cm]
      ${}^g$Collaborative Innovation Center of Quantum Matter, Beijing, China\\[0.3cm]
      ${}^h$Center for High Energy Physics, Peking University, Beijing 100871, China}
  \end{center}

\vspace{4ex}

\begin{abstract}
If  supersymmetry near the TeV scale is realized in Nature, the pair production of scalar top squarks is expected to be observable at the Large Hadron Collider. Recently, effective field-theory methods were employed to obtain approximate predictions for the cross section for this process, which include soft-gluon emission effects up to next-to-next-to-leading order (NNLO) in perturbation theory. In this work we employ the same techniques to resum soft-gluon emission effects to all orders in perturbation theory and with next-to-next-to-logarithmic (NNLL) accuracy. We analyze the effects of NNLL resummation on the stop-pair production cross section by obtaining NLO+NNLL predictions in pair invariant mass and one-particle inclusive kinematics. We compare the results of these calculations to the approximate NNLO predictions for the cross sections.
\end{abstract}

\end{titlepage}

\section{Introduction}

After the discovery of a Higgs boson in 2012, the search for supersymmetric partners of the Standard Model particles is one of the most important goals of the experimental program at the Large Hadron Collider (LHC). This is especially true in the light of the fact that the LHC is expected to run at a center-of-mass energy of 14\,TeV when operations will resume in 2015. Consequently, the LHC will be able to further investigate the existence of supersymmetric particles with masses in the TeV range. The Minimal Supersymmetric Standard Model (MSSM) with R-parity conservation predicts the production of supersymmetric particles in pairs. At a hadron collider such as the LHC, one expects to observe primarily the production of supersymmetric particles carrying color charge, such as squarks and gluinos. In unified supersymmetric theories the third generation of squarks can have masses which are significantly lighter than the masses of the first two generations of squarks, as a consequence of large Yukawa and soft couplings entering the evolution of the mass parameters from the unification scale down to low energies. Consequently, the lightest of the two supersymmetric partners of the top quark could be the lightest squark in the spectrum and the first supersymmetric particle to be observed at the LHC. 

This fact motivated the work of several groups, who in the last fifteen years obtained predictions for the top-squark pair production cross section with an accuracy beyond the leading order in supersymmetric quantum chromodynamics (SUSY-QCD). The calculation of the next-to-leading order (NLO) corrections to the production of top-squark pairs was carried out in \cite{Beenakker:1997ut}. The NLO corrections enhance the production cross section if the renormalization and factorization scales are chosen close to the top-squark mass. The NLO corrections to the stop production cross section, as well as the corrections to the production cross section for several other supersymmetric  particles, are implemented in the public codes {\tt Prospino} and {\tt Prospino~2} \cite{Beenakker:1996ed}. Electroweak corrections to stop pair production have a quite sizable effect on the tails of the invariant-mass and transverse-momentum distributions, but they only have a moderate impact on the total cross section. These corrections were evaluated in \cite{Hollik:2007wf,Beccaria:2008mi}. 

Corrections from soft gluon emissions account for a large fraction of the NLO SUSY-QCD corrections. For this reason the resummation of these corrections to next-to-leading logarithmic (NLL) accuracy was studied in \cite{Beenakker:2010nq} by means of a standard technique based upon the resummation of threshold logarithms in Mellin space. Within the same approach, the resummation  of the soft gluon corrections in the production of pairs of gluinos and  squarks of the first two generations was carried out up to next-to-next-to-leading logarithmic (NNLL) accuracy in \cite{Beenakker:2011sf,Beenakker:2013mva,Pfoh:2013iia}. Furthermore, in \cite{Langenfeld:2009eg,Langenfeld:2010vu,Langenfeld:2012ti} approximate next-to-next-to-leading order (NNLO) formulas for squark, stop, and gluino pair production were derived by means of resummation techniques.  These formulas include threshold corrections and Coulomb corrections. Recently, the fully differential NLO cross section for the production of squark pairs, including squark decays, was evaluated and matched with parton showers \cite{Gavin:2013kga}.

Over the last few years, an alternative approach to resummation, which makes use of soft-collinear effective theory (SCET) methods, has been developed \cite{Becher:2006nr,Becher:2006mr}. This approach, which allows one to work directly in momentum space, was applied to several processes of interest in collider physics, such as Drell-Yan production \cite{Becher:2007ty}, Higgs production \cite{Ahrens:2008nc,Ahrens:2008qu}, direct photon production \cite{Becher:2009th}, top-quark pair production \cite{Ahrens:2010zv,Ahrens:2011mw,Ferroglia:2012ku}, and slepton pair production \cite{Broggio:2011bd}. A similar method, combined with non-relativistic QCD techniques employed to resum Coulomb corrections, was independently developed in \cite{Beneke:2011mq} in order to study top-quark pair production at NNLL accuracy, and it was applied to the study of squark-pair, gluino-pair and stop-pair production at NLL accuracy \cite{Beneke:2010da,Falgari:2012hx}.

In particular, the studies of the top-pair differential distributions performed in \cite{Ahrens:2010zv,Ahrens:2011mw} can be repeated in a straightforward (but laborious) way for the production of top squarks. The method adopted in those works relies on the factorization of the partonic cross section, which takes place in the soft limit. The partonic cross section can in fact be expressed as the convolution of two different factors. Schematically, these factors are a hard function, which includes the effects of virtual corrections, and a soft function, which describes the emission of soft gluons from the external particles involved in the process. Two different kinds of soft limits were considered. In \cite{Ahrens:2010zv}, where the goal was the calculation of the invariant-mass distribution of the top-quark pair, the soft limit was defined as $z = M^2/s \to 1$, where $M$ is the pair invariant mass and $s$ is the square of the partonic center-of-mass energy. This framework is conventionally referred to as ``pair invariant mass'' (PIM) kinematics. In \cite{Ahrens:2011mw} instead, the goal was the calculation of the top-quark transverse-momentum and rapidity distributions, therefore the soft limit was defined as $s_4 \to 0$ with $s_4 = (p_4+k)^2 -m_t^2$, where $p_4$ is the momentum of the unobserved anti-top quark, $k$ is the momentum of the additional real radiation in the final state, and $m_t$ is the top-quark mass. The latter set up is know in the literature as ``one particle inclusive'' (1PI) kinematics. One can obtain predictions for the total cross section by integrating either one of the two kinds of distributions over the whole available phase space. The two calculations of the total cross section differ numerically because they neglect  different sets of terms that are formally subleading in the soft limit. Total cross-section predictions which account for this source of uncertainty are obtained by averaging results from calculations carried out in the two kinematic schemes \cite{Ahrens:2011px}. The calculations in PIM and 1PI kinematics share the same hard functions but require different soft functions. Furthermore, the soft functions in the two schemes do not depend on the spin of the particles involved in the process, so that the NLO soft functions employed in \cite{Ahrens:2010zv, Ahrens:2011mw} are the same soft functions that one needs in order to study top-squark pair production.

In \cite{Broggio:2013uba}, the hard functions for the production of top-squark pairs was evaluated up to NLO in SUSY-QCD. By combining the hard functions  with the soft functions of \cite{Ahrens:2010zv,Ahrens:2011mw}, it was possible to obtain approximate NNLO predictions for the pair invariant-mass and stop transverse-momentum distributions. At the moment, the most relevant observable in the study of top-squark production is the total cross section, which is employed in order to set lower bounds on the allowed values of the top-squark mass. For these reasons, in \cite{Broggio:2013uba} we integrated the differential distributions in order to obtain approximate NNLO formulas for the total top-squark production cross section. A systematic comparison showed that the approximate NNLO results of \cite{Broggio:2013uba} are compatible  with the NLL results of \cite{Beenakker:2010nq} and \cite{Falgari:2012hx}.

In this work we study the resummation of the soft-gluon corrections to the top-squark pair production cross section by solving the renormalization-group equations (RGEs) satisfied by the soft and hard functions. The known expressions for the relevant anomalous dimensions \cite{Ferroglia:2009ep,Ferroglia:2009ii} along with the NLO hard and soft functions calculated in \cite{Broggio:2013uba} and \cite{Ahrens:2010zv,Ahrens:2011mw}, respectively, are sufficient in order to carry out the resummation up to NNLL accuracy. The paper is organized as follows: In Section~\ref{sec:Notation} we summarize our notation and conventions. In Section~\ref{sec:Res} we review the resummation procedure, which uses the same scheme adopted in \cite{Ahrens:2010zv, Ahrens:2011mw} for the study of top-quark pair production. In Section~\ref{sec:Matching} we discuss the matching of the NNLL resummation considered here to fixed-order NLO calculations; the matching is carried out in order to obtain NLO+NNLL predictions for the total cross section. The phenomenological impact of these predictions and their relations to other studies found in the literature are presented in Section~\ref{sec:Pheno}. Finally, we collect our conclusions in Section~\ref{sec:Conc}.

\section{Notation}
\label{sec:Notation}

The production of top-squark pairs is described by the scattering process
\begin{align}
N_1(P_1) + N_2(P_2) &\to \st(p_3) + \st^*(p_4) + X(k) \, . \label{eq:hadproc}
\end{align}
We focus on the stop production at the LHC, so that $N_1$ and $N_2$ indicate the incoming protons, while $X$ is an inclusive hadronic final state. In this work, we treat the top squarks as on-shell particles and neglect their decay;  this approximation introduces an uncertainty of order $\Gamma_{\st}/m_{\st}$, where $m_{\st}$ is the stop mass and $\Gamma_{\st}$ represents its width.

The two partonic subprocesses contributing to the stop pair production at lowest order in perturbation theory are
\begin{align}\label{parton_processes}
q(p_1) + \bar{q}(p_2) &\to \st(p_3) + \st^*(p_4) \, , \nonumber \\
g(p_1) + g(p_2) &\to  \st(p_3) + \st^*(p_4) \, .
\end{align}
The momenta of the incoming partons $p_i\, (i=1,2)$ are related to the hadronic momenta through the relation $p_i = x_i P_i$. The relevant invariants for the hadronic scattering process are 
\begin{align}
S = (P_1+P_2)^2 \, ,  \quad T_1 = (P_1 -p_3)^2 - m_{\st}^2 \, , \quad U_1 = (P_1 - p_4)^2 - m_{\st}^2 \, .
\end{align}
In order to describe the partonic scattering, we employ the Mandelstam invariants
\begin{align}
s = x_1 x_2 S = (p_1 +p_2)^2 \, , \quad t_1 = x_1 T_1 \, , \quad u_1 = x_2 U_1 \, , \nonumber \\
M^2 = (p_3 + p_4)^2 \, , \quad  s_4 = s+t_1 +u_1 = (p_4+k)^2 - m_{\st}^2 \, . \label{eq:ManPart}
\end{align}
In Born approximation $s+t_1+u_1=0$, and consequently $M^2 = s$ and $s_4 = 0$.

Following the procedure employed in \cite{Broggio:2013uba} and in the papers devoted to the calculation of differential distributions for top-quark pair production \cite{Kidonakis:2001nj,Ahrens:2009uz,Ahrens:2010zv,Ahrens:2011mw}, we consider two different kinematic schemes, each of which has its own threshold limit. In PIM kinematics the threshold region is defined by the limit $s \to M^2$, while in 1PI kinematics the threshold region is approached by taking the limit $s_4 \to 0$. The two different kinematics are suitable for the calculation of different differential distributions: PIM kinematics is used in order to calculate the pair invariant-mass distribution, while 1PI kinematics is employed in order to evaluate the stop transverse-momentum and rapidity distributions. In contrast to the production threshold region, which is defined by the limit $\beta = \sqrt{1- 4 m^2_{\st}/s } \to 0$ and is often employed in the calculation of the total cross section in the soft limit,  in the PIM and 1PI threshold regions top squarks are not necessarily produced nearly at rest. For instance, if we require to observe a stop pair with an invariant mass $M$, the squared partonic center-of-mass energy should be larger than $M^2$, which can be much larger than the production threshold $s\ge 4 m_{\st}^2$. In both kinematic schemes, the partonic cross section in the threshold region is numerically dominated by the contribution of soft gluon emission. 

A fact which is particularly relevant for resummation purposes is that in the soft limit the partonic cross section factors into products of hard and soft functions. Each of these two factors  satisfies known RGEs. The anomalous dimensions entering these equations are know up to NNLO \cite{Ferroglia:2009ep,Ferroglia:2009ii}, while the matching coefficients are known up to NLO. This allows one to solve the RGE in Laplace space \cite{Becher:2006nr,Becher:2006mr} and obtain resummed formulas which are valid up to NNLL accuracy. 

\subsection{PIM kinematics}

In order to deal with PIM kinematics, it is useful to introduce the quantities
\be
z = \frac{M^2}{s} \, , \qquad \tau = \frac{M^2}{S} \, , \qquad \beta_{\st} = \sqrt{1- \frac{4 m_{\st}^2}{M^2}} \, ,
\ee
the threshold region is defined by the limit $z \to 1$. Because of the QCD factorization theorem \cite{Collins:1989gx}, the double-differential cross section  in $M$  and $\theta$ (the stop scattering angle in the partonic rest frame) can be factorized as
\be
\frac{d^2 \sigma}{d M d \cos{\theta}} = \frac{\pi \beta_{\st}}{S M} \sum_{i,j} \int_\tau^1 \frac{dz}{z} \ff_{ij}\left(\frac{\tau}{z}, \mu_f\right) C_{{\rm PIM}, ij}\left(z,M, \cos{\theta}, \mu_f\right) ,
\label{eq:dMdcos}
\ee
where $\mu_f$ is the factorization scale, and the sum runs over the incoming partons.\footnote{In the following we drop the subscript PIM (and the corresponding subscript 1PI) whenever this does not lead to ambiguities.}%
As usual, parton luminosities $\ff_{ij}$ are defined as the convolutions of the non-perturbative parton distribution functions (PDFs) for the incoming partons:
\be
\ff_{ij}(y,\mu_f) = \int_y^1 \frac{dx}{x} f_{i/N_1}\left(x,\mu_f\right) f_{j/N_2} \left(\frac{y}{x}, \mu_f\right) \equiv f_{i/N_1}(y)\otimes f_{j/N_2}(y)\, .
\ee
The functions $C_{ij}$ in Eq.~(\ref{eq:dMdcos}) are the hard-scattering kernels, which are related to the partonic cross sections and can be calculated in perturbation theory. The hard-scattering kernels depend on the top-squark masses $m_{\st}$ and $m_{\tilde{t}_2}$ (where we assume $m_{\st} < m_{\tilde{t}_2}$), the mass $m_{\tilde{q}}$ of the first two generations of squarks and of the sbottoms (which we assume to be all degenerate), the top-quark mass $m_t$, the gluino mass $m_{\tilde{g}}$, and the $\tilde{t}_1$-$\tilde{t}_2$ mixing angle $\alpha$. However, in order to avoid the use of  an unnecessarily heavy notation, we drop these quantities from the list of arguments of the hard-scattering kernels.

At lowest order in $\alpha_s$, only the quark annihilation and gluon fusion channels contribute to the hard-scattering kernels, therefore $ij \in \{q \bar{q}, gg\}$. In order to go beyond leading order, one needs to consider virtual and real emission corrections to the Born approximation, so that new production channels such as $q g \to \st \st^* q$ open up.  However, it is a well-known fact that both the hard-gluon emission and the additional production channels are suppressed by powers of $(1-z)$ and can be safely neglected while deriving results within the partonic-threshold limit. Therefore, Eq.~(\ref{eq:dMdcos}) can be rewritten as
\begin{align}
\frac{d^2 \sigma}{d M d \cos{\theta}} &= \frac{\pi \beta_{\st}}{S M} \int_\tau^1 \frac{dz}{z} \Bigl[ \ff_{gg}\left(\frac{\tau}{z},\mu_f \right)
C_{gg}\left(z,M, \cos{\theta},\mu_f \right) \nonumber \\
&  + \ff_{q \bar{q}} \left(\frac{\tau}{z},\mu_f \right) C_{q \bar{q}}\left(z,M, \cos{\theta},\mu_f \right) + 
\ff_{\bar{q} q} \left(\frac{\tau}{z},\mu_f \right) C_{q\bar{q}}\left(z,M,-\cos{\theta},\mu_f \right)  \Bigr] + \dots \, ,
\label{eq:abx}
\end{align}
where we omit terms of ${\cal O}(1-z)$. In Eq.~(\ref{eq:abx})  the quark channel luminosities $\ff_{q \bar{q}}$ and $\ff_{\bar{q} q}$ are understood to be summed over all light quark flavors. The two terms in the second line of Eq.~(\ref{eq:abx}) differ in the fact that in the first term the quark (antiquark) comes from the hadron $N_1$ ($N_2$) in Eq.~(\ref{eq:hadproc}), while in the second term the quark (antiquark) comes from the hadron $N_2$ ($N_1$), respectively. The total cross section can be obtained by integrating over $\cos \theta$ in the range $[-1,1]$ and over $M$ in the range $[2 m_{\st}, \sqrt{S}]$.

In the soft limit $z \to 1$, the hard-scattering kernels $C_{ij}$ factor into a product of hard and soft functions \cite{Ahrens:2010zv}:
\be
C_{ij}(z,M,\cos{\theta},\mu_f) = \Tr\left[\bm{H}_{ij}(M,\cos{\theta},\mu_f) 
\bm{S}_{ij}(\sqrt{s}(1-z),M,\cos{\theta},\mu_f)\right] + 
{\mathcal O}(1-z) \, . \label{eq:Fact}
\ee
Here and in what follows we employ boldface fonts to indicate matrices in color space, such as the hard functions $\bm{H}_{ij}$ and the soft functions $\bm{S}_{ij}$.\footnote{Following the notation adopted in \cite{Broggio:2013uba}, we drop the top-quark mass and the SUSY parameters from the  arguments of the hard functions as well as the stop mass from the arguments of the soft functions.}%
Throughout this paper, we work in the $s$-channel singlet-octet basis already employed in \cite{Broggio:2013uba}.

A factorization formula analogous to  Eq.~(\ref{eq:Fact}) for the top-quark pair production was derived  by employing SCET and heavy-quark effective theory in \cite{Ahrens:2010zv}. A completely analogous procedure can be followed in order to derive Eq.~(\ref{eq:Fact}), which is valid in the case of top-squark pair production. The hard functions, computed in \cite{Broggio:2013uba}, are obtained from virtual corrections and are ordinary functions of their arguments. The soft functions arise from the real emission of soft gluons and contain distributions which are singular in the $z \to 1$ limit. The soft functions are identical to the ones needed for the case of top-quark pair production, which were evaluated up to NLO in \cite{Ahrens:2010zv}. The hard functions were evaluated up to NLO in \cite{Broggio:2013uba}. The RGEs satisfied by the hard and soft functions are identical to the ones satisfied by the corresponding quantities in the top quark production case and are discussed in detail in \cite{Ahrens:2010zv}. The anomalous dimensions regulating these RGEs are know up to NNLO. As discussed in Section~\ref{sec:Res}, by solving these RGEs it is possible to implement the resummation of soft gluon emission corrections up to NNLL accuracy.

\subsection{1PI kinematics}

1PI kinematics is used whenever one needs to consider kinetic properties of a single particle, rather than of the pair. One can then write the double-differential distribution in the top-squark transverse momentum and rapidity as 
\be
\frac{d^2 \sigma}{d p_T d y} = \frac{2 \pi p_T}{S} \sum_{ij} \int_{x_{1}^{{\rm min}}}^1 \frac{d x_1}{x_1}  \int_{x_{2}^{{\rm min}}}^1 \frac{d x_2}{x_2}
f_{i/N_1}(x_1,\mu_f) f_{j/N_2}(x_2,\mu_f)\,C_{{\rm 1PI}, ij}\left(s_4,s,t_1,u_1,\mu_f\right) .
\label{eq:1PIdiff}
\ee
Obviously, only the quark-annihilation and gluon-fusion channels contribute to the hard-scattering kernels $C_{ij}$ at the lowest order in $\alpha_s$. The hadronic Mandelstam variables $T_1$ and $U_1$ can be expressed in terms of   the stop rapidity and transverse momentum as
\be
T_1 = -\sqrt{S} m_\perp e^{-y} \, , \qquad U_1 = -\sqrt{S} m_\perp e^y \, ,
\ee
where $m_\perp = \sqrt{p_T^2 + m_{\st}^2}$. Therefore, the variables $s,s_4,t_1,u_1$, which are arguments of the 1PI scattering kernels, can be expressed in terms of $p_T,y,x_1,x_2$. The lower integration limits in Eq.~(\ref{eq:1PIdiff}) are
\be
x_1^{{\rm min}} = - \frac{U_1}{S+T_1} \, , \qquad x_2^{{\rm min}} = - \frac{x_1 T_1}{x_1 S + U_1} \, .
\ee
In order to obtain the total cross section, it is necessary to integrate the double-differential distribution in Eq.~(\ref{eq:1PIdiff}) with respect to the top-squark rapidity and transverse momentum over the range
\be
0 \le |y| \le \frac{1}{2} \ln{\frac{1 + \sqrt{1-4 m_\perp^2/S}}{1- \sqrt{1-4 m_\perp^2/S}}} \, , \qquad 0 \le p_T \le \sqrt{\frac{S}{4} - m_{\st}^2} \, .
\ee

In the case of 1PI kinematics, the hard-scattering kernels in the soft limit $s_4 \to 0$ factor into a product of hard and soft functions, in analogy to Eq.~(\ref{eq:Fact}):
\be\label{soft1PI}
C_{ij}(s_4,s',t'_1,u'_1,\mu) = \Tr \left[\bm{H}_{ij}(s',t'_1,u'_1,\mu) 
\bm{S}_{ij}(s_4,s',t'_1,u'_1,\mu) \right]  + {\mathcal O}(s_4) \, .
\ee
As emphasized in \cite{Ahrens:2011mw}, the Mandelstam invariants $s',t'_1,u'_1$ can differ from $s, t_1,u_1$ by power corrections proportional to $s_4$. For example, explicit results for the hard and soft functions can be rewritten by employing either the relation $s'+t'_1+u'_1 = 0$ or $s'+t'_1+u'_1 = s_4$. The difference between the two choices is due to terms suppressed by positive powers of $s_4$. We deal with this ambiguity following the methods described in Section~4 of \cite{Ahrens:2011mw}.

As in the case of PIM kinematics, the hard and soft functions are matrices in color space, arising  from virtual and soft-emission corrections, respectively. The 1PI hard functions are identical to the ones encountered in the PIM kinematics. The 1PI soft functions, which differ from those derived in PIM kinematics, depend on plus distributions which are singular in the limit $s_4 \to 0$. They were originally computed up to NLO in \cite{Ahrens:2011mw} for the top-quark pair production cross section. The RGEs satisfied by the hard and soft functions are identical to the ones discussed in \cite{Ahrens:2011mw}, therefore all of the elements are in place to implement the resummation up to NNLL accuracy.

\section{Resummation}
\label{sec:Res}

Our main goal is to resum the leading singular terms in $(1-z)$ (PIM kinematics) or $s_4$ (1P1 kinematics) in the region of (partonic) phase-space where the stop production cross section is dominated by the threshold terms. This is accomplished by deriving and solving RGEs for the hard and soft functions. The RGEs for the hard functions do not depend on the virtual particles running in the loops or on the spin of the final state particles, and therefore they are precisely the same equations that have been discussed and solved in \cite{Ahrens:2010zv,Ahrens:2011mw} up to the order appropriate for NNLL resummation. The RGE satisfied by the PIM soft functions and its solution can be found in Section~5.1 of \cite{Ahrens:2010zv}, while the solution of the RGE satisfied by the 1PI soft functions can be found in Section~3.2 of \cite{Ahrens:2011mw}. Here we limit ourselves to collect the resummation formulas for the hard-scattering kernels appearing in Eqs.~(\ref{eq:dMdcos}) and (\ref{eq:1PIdiff}).

The resummed expression for the hard-scattering kernels in PIM kinematics is 
\begin{align}
  \label{eq:MasterFormulaPIM}
  C(z,M,\cos\theta,\mu_f) &= \exp \big[ 4a_{\gamma^{\phi}}(\mu_s,\mu_f) \big]
  \nonumber
  \\
  &\hspace{-2em} \times \Tr \Bigg[ \bm{U}(M,\cos\theta,\mu_h,\mu_s) \,
  \bm{H}(M,\cos\theta,\mu_h) \, \bm{U}^\dagger(M,\cos \theta,\mu_h,\mu_s)
  \nonumber
  \\
  &\hspace{-2em} \times \tilde{\bm{s}}
  \left(\ln\frac{M^2}{\mu_s^2}+\partial_\eta,M,\cos\theta,\mu_s\right) \Bigg]
  \frac{e^{-2\gamma_E \eta}}{\Gamma(2\eta)} \frac{z^{-\eta}}{(1-z)^{1-2\eta}} \, ,
\end{align}
where we dropped the indices indicating the partonic channel. The channel-dependent hard matrices $\bm{H}$ are described in Section~3.1 of \cite{Broggio:2013uba}, where they were evaluated up to NLO. The Laplace transform of the soft matrices, $\tilde{\bm{s}}$, was defined in Section~4.2 of \cite{Ahrens:2010zv}. The introduction of the Laplace transform of the soft matrices is motivated by the fact that, in Laplace space, soft functions are regular polynomials of their first argument, which satisfy  ordinary first-order differential equations \cite{Becher:2006nr}. The PIM evolution matrices $\bm{U}$ and the exponential factor $a_{\gamma^\phi}$ are defined in Section~5 of \cite{Ahrens:2010zv}. The parameter $\eta$ arises from the solution of the RGE for the Laplace-transformed soft function $\tilde{\bm{s}}$. The notation is such that one must first take the derivatives with respect to $\eta$ appearing in the first argument of $\tilde{\bm{s}}$ and then set $\eta = 2 a_\Gamma(\mu_s,\mu_f)$, as discussed in Section~5 of \cite{Ahrens:2010zv}. For values $\mu_s < \mu_f$ one finds that $\eta <0$ and consequently one must use a subtraction at $z=1$ and analytic continuation to express integrals in terms of plus distributions \cite{Bosch:2004th}. For example, for a smooth function $g(z)$ that is not singular for  $z \to 1$ one can analytically continue the integrals from the region $\eta > 0$ to the region $\eta > -1/2$ by means of the relation
\begin{align}
\int_{\tau}^1 dz \frac{g(z)}{(1-z)^{1-2 \eta}} & =\int_{\tau}^1 dz\,  \frac{g(z) - g(1)}{(1-z)^{1-2 \eta}} + \frac{g(1)}{2 \eta} (1-\tau)^{2 \eta} \, . \label{eq:subz}
\end{align} 
If necessary, it is possible to analytically continue the integral on the left-hand side of this equation to the region $\eta > - n/2$ for an arbitrary positive integer $n$. This can be done by subtracting an increasing number of terms from the Taylor expansion of $g(z)$ at $z=1$.

Although the all-order hard-scattering coefficients $C$ depend on the factorization scale $\mu_f$ but do not depend  on the soft and hard scales $\mu_s$ and $\mu_h$, any practical implementation of the resummation formula Eq.~(\ref{eq:MasterFormulaPIM}) will have a residual dependence on these two scales. This is due  to the fact that  the anomalous dimensions appearing in the evolution factors in Eq.~(\ref{eq:MasterFormulaPIM}) are evaluated up to a given finite order in perturbation theory. The order at which this truncation takes places, together with the order at which the hard and soft functions are evaluated, defines the accuracy at which the resummation formula is implemented. The anomalous dimensions and the hard and soft functions are known at an order which is sufficient to carry out the resummation with NNLL accuracy. The choice of the numerical values for the hard and soft scales is discussed in Section~\ref{sec:Matching}.

The resummation formula for the hard-scattering kernels in 1PI kinematics is (see Section~3.2 in \cite{Ahrens:2011mw})
\begin{align}\label{eq:MasterFormula1PI}
 C(s_4,s',t'_1,u'_1,\mu_f) &= \exp \Bigg[ 2 a_\Gamma(\mu_s,\mu_f) \ln{\frac{m_{\st}^2 \mu_s'^2}{t'_1 u'_1}}+ 4a_{\gamma^{\phi}}(\mu_s,\mu_f) \Bigg]
  \nonumber
  \\
  &\hspace{-2em} \times \Tr \Bigg[ \bm{U}(s',t'_1,u'_1,\mu_h,\mu_s) \,
  \bm{H}(s',t'_1,u'_1,\mu_h) \, \bm{U}^\dagger(s',t'_1,u'_1,\mu_h,\mu_s)
  \nonumber
  \\
  &\hspace{-2em} \times \tilde{\bm{s}}
  \left(\partial_\eta,s',t'_1,u'_1,\mu_s\right) \Bigg]
  \frac{e^{-2\gamma_E \eta}}{\Gamma(2\eta)} \frac{1}{s_4} \left(\frac{s_4}{\sqrt{s_4 +m^2_{\st}} \mu_s}\right)^{2 \eta} \, .
\end{align}
The evolution factors and the hard functions in Eq.~(\ref{eq:MasterFormula1PI}) are the same as in the PIM case (see \cite{Ahrens:2010zv}). The Laplace transform of the  1PI soft function $\tilde{\bm{s}}$ was evaluated up to NLO and can be found in Section~3.1 of \cite{Ahrens:2011mw}. As for the PIM case, for values of the scale such that $\eta < 0$ one must use analytic continuation to interpret the formula in terms of plus distributions. Also in the case of 1PI kinematics, the resummation of the top-squark pair production cross section can be carried out at NNLL accuracy.

\section{Matching and scale choices}
\label{sec:Matching}

Although the method employed allows us to obtain predictions for the pair invariant-mass distribution of the stop pair and for the transverse-momentum and rapidity distribution of a single top squark, we will limit ourselves to the calculation of the observable of phenomenological interest at the moment, i.e.\ the total stop-pair production cross section. The total cross section can be obtained by integrating the double-differential distributions in PIM and 1PI kinematics over the complete phase space, as explained in Section~\ref{sec:Notation}. 

Obviously, one wants to combine  NNLL resummation with the most accurate fixed-order calculations of the total cross section available to date. Currently, the total stop-pair production cross section is known at NLO \cite{Beenakker:1997ut}. The NLO calculations can be matched to NNLL calculations of the total cross section as follows:
\begin{align} \label{eq:matching}
\sigma_i^{\tNLO + \tNNLL} &\equiv \left.\sigma_i^{\tNNLL}\right|_{\mu_h, \mu_s, \mu_f} + \left.\sigma_i^{\tNLO , \mbox{{\tiny subleading}}} \right|_{\mu_f} \, , \nonumber \\
&\equiv \left.\sigma_i^{\tNNLL}\right|_{\mu_h, \mu_s, \mu_f}  + \left( \left. \sigma_i^{\tNLO } \right|_{\mu_f} - \left.\sigma_i^{\tNLO,  \mbox{{\tiny leading}}}\right|_{\mu_f} \right) ,
\end{align}
where the subscript $i \in \{ \mbox{PIM}, \mbox{1PI}\}$ indicates the kinematic scheme employed. Furthermore, the subscripts in Eq.~(\ref{eq:matching}) indicate the scales ($\mu_f,\mu_h,\mu_s$) on which each term depends. In Eq.~(\ref{eq:matching}), $\sigma_i^{\tNLO }$ is the exact result in fixed-order perturbation theory, while 
\begin{align}
\sigma_i^{\tNLO , \mbox{{\tiny leading}}}  &\equiv \left.\sigma_i^{\tNNLL}\right|_{\mu_h = \mu_s = \mu_f} \, 
\end{align}
captures the leading singular terms in the threshold limit. If the various scales are set equal to each other, the resummed expressions for the cross section automatically reduce to fixed-order perturbative expansions. Consequently, the second term in the first line of Eq.~(\ref{eq:matching}) includes the set of NLO terms which are not included in the resummed formulas, and it can be added to the first term, which includes the NNLL corrections, without introducing any double counting. The issue of the choice of numerical default value for the scales in the first term on the right-hand side of Eq.~(\ref{eq:matching}) is addressed below. NLO predictions for the stop pair cross section can be conveniently obtained from the programs {\tt{Prospino}} and {\tt{Prospino2}} \cite{Beenakker:1996ed}. The matching procedure of Eq.~(\ref{eq:matching}) can be carried out separately for each of the two kinematic schemes considered.

Since the total cross section can be obtained starting from either of the two kinematics, but each kinematics neglects different sets of subleading corrections, we follow the procedure already adopted in \cite{Ahrens:2011px,Broggio:2013uba} and average the two results. Schematically, our resummed prediction for the total cross section is then obtained as 
\begin{align}
\sigma_{\tNLO + \tNNLL} = \frac{1}{2} \left(\sigma^{\mbox{{\tiny PIM}}}_{\tNLO + \tNNLL} + \sigma^{\mbox{{\tiny 1PI}}}_{\tNLO + \tNNLL}  \right) . \label{eq:average}
\end{align} 
Similarly, in evaluating the perturbative error associated with our result, we want to reflect also the uncertainty associated to the choice of the kinematic scheme. In order to achieve this goal, we start by varying separately each scale $\mu_i$ ($i=f,h,s$) in the range $[\mu_{0,i}/2, 2 \mu_{0,i}]$, where $\mu_{0,i}$ denotes the default choice for the scale $\mu_i$, which is discussed in the next two sections. We then evaluate the quantities
\begin{align}
\Delta \sigma_f^{+} &\equiv \mbox{max}\Bigl\{\sigma^{\mbox{{\tiny PIM}}} (\mu_{0,f},\mu_{0,h},\mu_{0,s}), \sigma^{\mbox{{\tiny PIM}}} (2 \mu_{0,f},\mu_{0,h},\mu_{0,s}),  \sigma^{\mbox{{\tiny PIM}}} (\mu_{0,f}/2,\mu_{0,h},\mu_{0,s}),
\nonumber  \\
&\qquad \sigma^{\mbox{{\tiny 1PI}}} (\mu_{0,f},\mu_{0,h},\mu_{0,s}), \sigma^{\mbox{{\tiny 1PI}}}(2 \mu_{0,f},\mu_{0,h},\mu_{0,s}),   \sigma^{\mbox{{\tiny 1PI}}} (\mu_{0,f}/2,\mu_{0,h},\mu_{0,s})  \Bigr\} - \frac{\sigma^{\mbox{{\tiny PIM}}} + \sigma^{\mbox{{\tiny 1PI}}}}{2} \, , \nonumber \\
\Delta \sigma_f^{-} &\equiv \mbox{min}\Bigl\{\sigma^{\mbox{{\tiny PIM}}} (\mu_{0,f},\mu_{0,h},\mu_{0,s}), \sigma^{\mbox{{\tiny PIM}}} (2 \mu_{0,f},\mu_{0,h},\mu_{0,s}),  \sigma^{\mbox{{\tiny PIM}}} (\mu_{0,f}/2,\mu_{0,h},\mu_{0,s}),
\nonumber  \\
&\qquad \sigma^{\mbox{{\tiny 1PI}}} (\mu_{0,f},\mu_{0,h},\mu_{0,s}), \sigma^{\mbox{{\tiny 1PI}}}(2 \mu_{0,f},\mu_{0,h},\mu_{0,s}),   \sigma^{\mbox{{\tiny 1PI}}} (\mu_{0,f}/2,\mu_{0,h},\mu_{0,s})  \Bigr\} - \frac{\sigma^{\mbox{{\tiny PIM}}} + \sigma^{\mbox{{\tiny 1PI}}}}{2}\, , \label{eq:sigmas}
\end{align}
where we neglected the subscript NLO $+$ NNLL for each of the cross sections appearing on the left-hand side of Eqs.~(\ref{eq:sigmas}). In complete analogy, we also evaluate the quantities $\Delta \sigma^\pm_h$ and $\Delta \sigma^\pm_s$ by varying the hard or soft scales, while keeping the other two scales equal to their default values. Finally, the perturbative uncertainty on the cross section is obtained by combining the quantities $\Delta \sigma_i^{\pm}$ in quadrature, i.e.\
\begin{align}
\Delta \sigma_\mu^\pm &\equiv \sqrt{ \left(\Delta \sigma_f^\pm\right)^2 +\left(\Delta \sigma_h^\pm\right)^2+ \left(\Delta \sigma_s^\pm\right)^2  }  \, .
\end{align}

At this stage we turn our attention to the choice of the default values for the soft, hard and factorization scales.

\subsection{Choice of the hard and factorization scales \label{sec:muh}}

The hard scale $\mu_h$ should be set to the characteristic scale of the underlying partonic subprocesses shown in Eq.~(\ref{parton_processes}). An obvious possibility would be the invariant mass $M$ of the stop pair, which is the lower bound on the partonic center-of-mass energy $\sqrt{s}$. However, the observable $M$ is only defined in PIM kinematics, whereas the pair invariant mass is not observed in 1PI kinematics. We will therefore use the other obvious possibility, the production threshold $\mu_{0,h}=2 m_{\st}$, as the default value for the hard scale in both kinematic schemes. For the factorization scale, we follow the standard choice made in fixed-order perturbation theory calculations, namely we set $\mu_{0,f}=m_{\st}$. As is common practice, we will vary the scales $\mu_h$ and $\mu_f$ independently by factors of 2 about the default values.

\subsection{Choice of the soft scale \label{sec:mus}}

\begin{figure}[t]
\begin{center}
\includegraphics[width=0.48\textwidth]{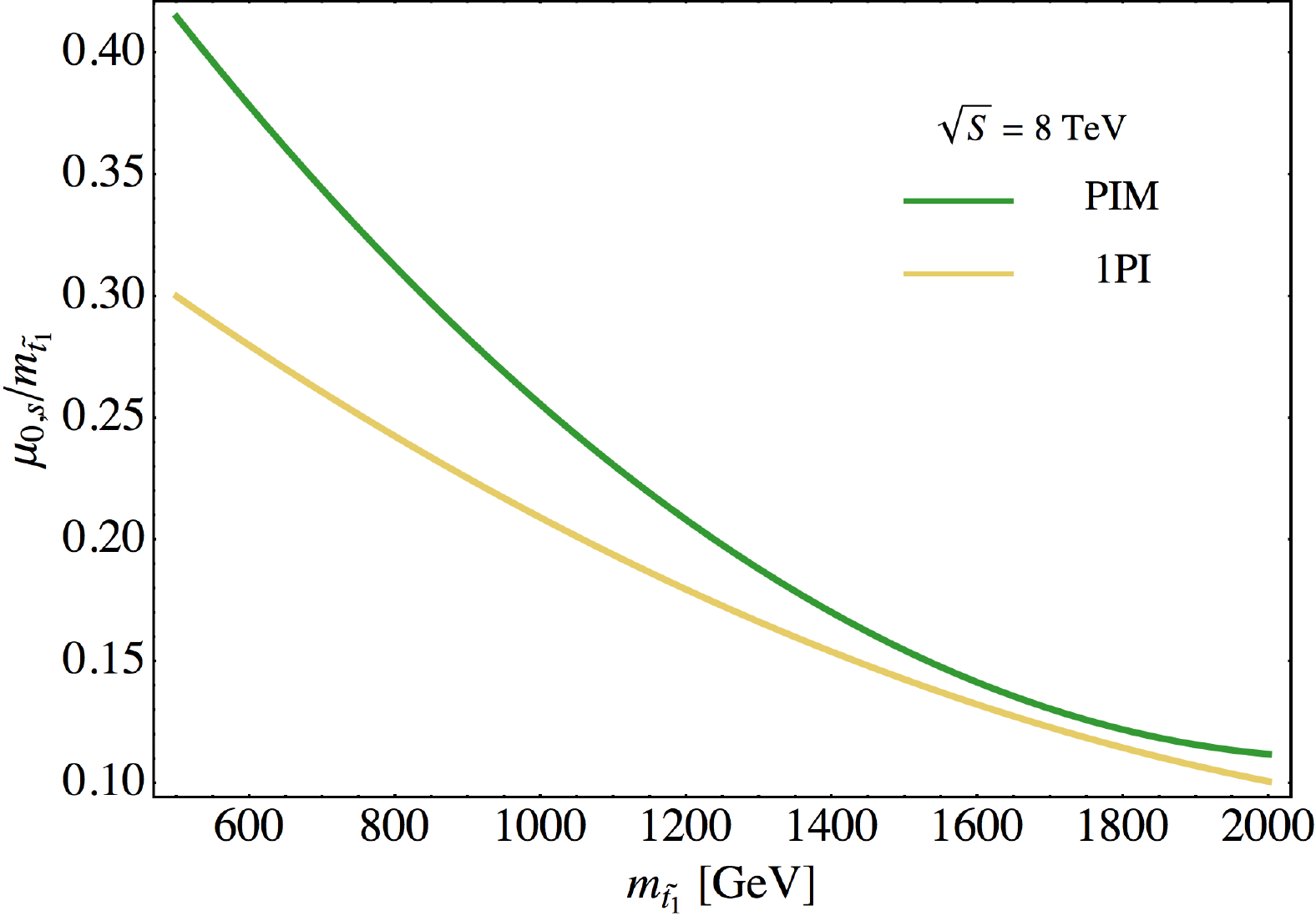}
\end{center}
\caption{\label{fig:muschoice} 
Dependence of the default value $\mu_{0,s}$ for the soft scale (in units of $m_{\st}$) on the top-squark mass, for PIM (green line) and 1PI kinematics (orange line) kinematics. The plot refers to the LHC operating at a center-of-mass energy of $\sqrt{S}=8$\,TeV.}
\end{figure}

Contrary to the hard matching scale, the soft matching scale is not associated with a parameter entering the partonic cross sections. Rather, it is generated dynamically when the partonic cross sections are convoluted with the steeply falling PDFs \cite{Becher:2006mr}. Our procedure for fixing the value of the soft scale is similar to the one employed in the case of top-quark pair production in \cite{Ahrens:2010zv,Ahrens:2011mw}. In the case of the top-squark pair production considered here, the problem is slightly more complicated because  the stop mass is not known, and it becomes a parameter in the determination of $\mu_s$. In general, one expects to find  that the soft function has a well-behaved perturbative expansion when $\mu_s$ is set equal to a scale characteristic of the energy of the real soft radiation, which is expected to be smaller than the hard scales $m_{\st}$ and $\sqrt{s}$. In order to find this scale for a given kinematic scheme and fixed center-of-mass energy and $m_{\st}$, we look for the minimum of the $\alpha_s$ corrections to the total cross section arising from the soft function as a function of $\mu_s$. In order to isolate these corrections, we select the part of the NNLL resummed formula for the hard-scattering kernels which arises from $\tilde{\bm{s}}^{(1)}$ (i.~e. the NLO contribution to the soft function), evaluate the contribution of these terms  to the total cross section, and divide what we find by the NLL cross section. We furthermore set $\mu_s = \mu_f = \mu_h$, which is equivalent to considering the fixed-order corrections at NLO accuracy. When plotting these corrections as a function of $\mu_s/m_{\st}$ for fixed $s$ and $m_{\st}$, one finds that they show a minimum. We further plot the location of the minimum as a function of $m_{\st}$. The curve which emerges is that of a smooth, monotonically decreasing function, which for fixed kinematics and collider energy can be well approximated by a quadratic polynomial. We employ such fits in order to determine the default value of the soft scale for fixed $m_{\st}$ and $S$. For example, for $\sqrt{S} = 8$\,TeV, $m_{\st} \in [500,2000]$\,GeV, and assuming PIM kinematics, we fix the soft scale using the formula (with $m_{\st}$ in GeV)
\be
\mu_{0,s} = m_{\st} \left( 0.632 - 4.93 \times 10^{-4}\, m_{\st} 
+ 1.17 \times 10^{-7}\, m_{\st}^2 \right) .
\ee
A similar curve is found in the case of 1PI kinematics. The resulting functions are shown in Figure~\ref{fig:muschoice}. In order to account for the uncertainty introduced by the scale choice, in phenomenological predictions we allow the chosen soft scale to vary in the range $[\mu_{0,s}/2, 2 \mu_{0,s}]$, as explained above.

\section{Phenomenology}
\label{sec:Pheno}

\begin{table}
\begin{center}
\begin{tabular}{|c|c|c|c|}
\hline $m_{\tilde{g}}$ & 1489.98\,GeV & $m_{\tilde{t}_2}$ & 1319.87\,GeV \\ 
\hline $m_t$ & 173.3\,GeV & $m_{\tilde{q}}$ & 1460.3\,GeV \\ 
\hline$\alpha$ & $68.4^\circ$ & & \\ 
\hline 
\end{tabular}
\end{center}
\caption{\label{tab:inputs} 
SUSY parameters other than $m_{\st}$ characterizing the benchmark point {\tt 40.2.5} in \cite{AbdusSalam:2011fc}.}
\end{table} 

In this section we analyze the numerical predictions for the stop pair production cross section at NLO+NNLL accuracy. In particular, {\em i)} we compare the results obtained in PIM and 1PI kinematics and their average, {\em ii)}  we investigate the dependence of the predictions on the variation of the hard, soft and factorization scales, {\em iii)} we provide numerical tables for different values of the stop mass and for different choices of the PDF sets, and {\em iv)} we compare the predictions with NLO+NNLL accuracy to the approximate NNLO cross section studied in \cite{Broggio:2013uba}. In order to keep the presentation concise, we consider two values for the LHC center-of-mass energy: $\sqrt{S} = 8$\,TeV, which is the energy at which the machine was running before the shutdown in 2013-2014, and $\sqrt{S} = 14$\,TeV, which is the targeted energy when operations resume in 2015. Furthermore, as in \cite{Broggio:2013uba} we fix the SUSY parameters other than the light stop mass to the value characterizing the benchmark point {\tt 40.2.5} in \cite{AbdusSalam:2011fc}. As it was shown in \cite{Broggio:2013uba}, the total cross section shows little sensitivity to the SUSY parameters other than $m_{\st}$. Table~\ref{tab:inputs} collects the values of the input parameters entering the hard functions employed throughout this section. The benchmark point {\tt 40.2.5} uses a value of $1087.15$\,GeV for the top-squark mass. We employ this value in the tables below.\footnote{Readers interested in predictions for other values of the stop mass (or different input parameters) can contact the authors.}%
However, in the same tables we also consider $m_{\st} = 500$\,GeV, which is representative of the current experimental lower bounds on this quantity. In addition, we plot mass scans for the total cross section in the range $m_{\st} \in [500,2000]$\,GeV. In the following, unless we explicitly write that we do otherwise, it is understood that we employ NNLO PDFs in NLO+NNLL calculations and approximate NNLO calculations, while we employ NLO PDFs in NLO and NLL calculations. In each plot or table, we explicitly indicate the use of either CT10 \cite{Lai:2010vv,Gao:2013xoa} or MSTW2008 \cite{Martin:2009iq} PDFs.

\subsection{Comparison between 1PI and PIM kinematics}

\begin{figure}[t]
\begin{center}
\begin{tabular}{cc}
\includegraphics[width=0.48\textwidth]{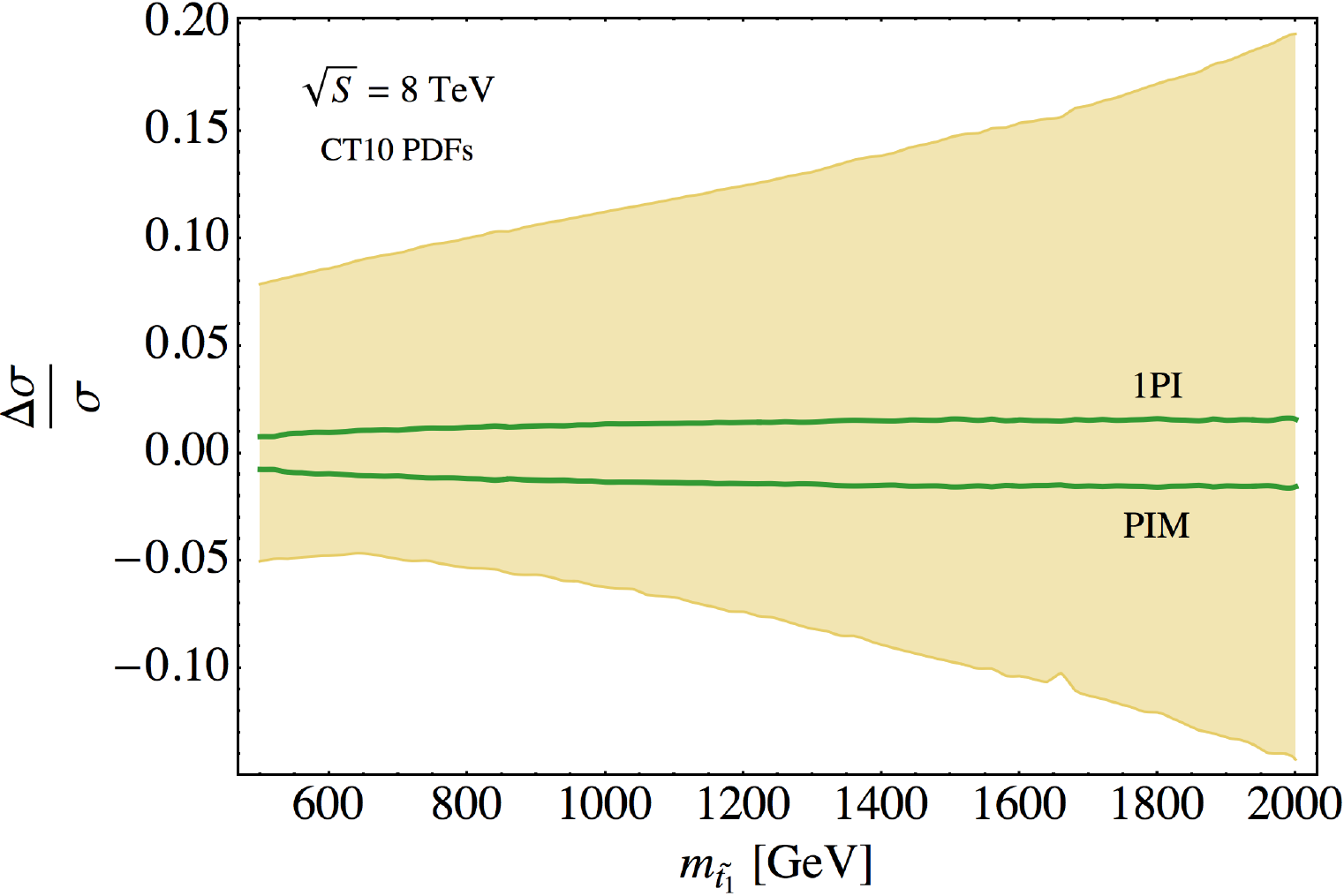}  &  \includegraphics[width=0.48\textwidth]{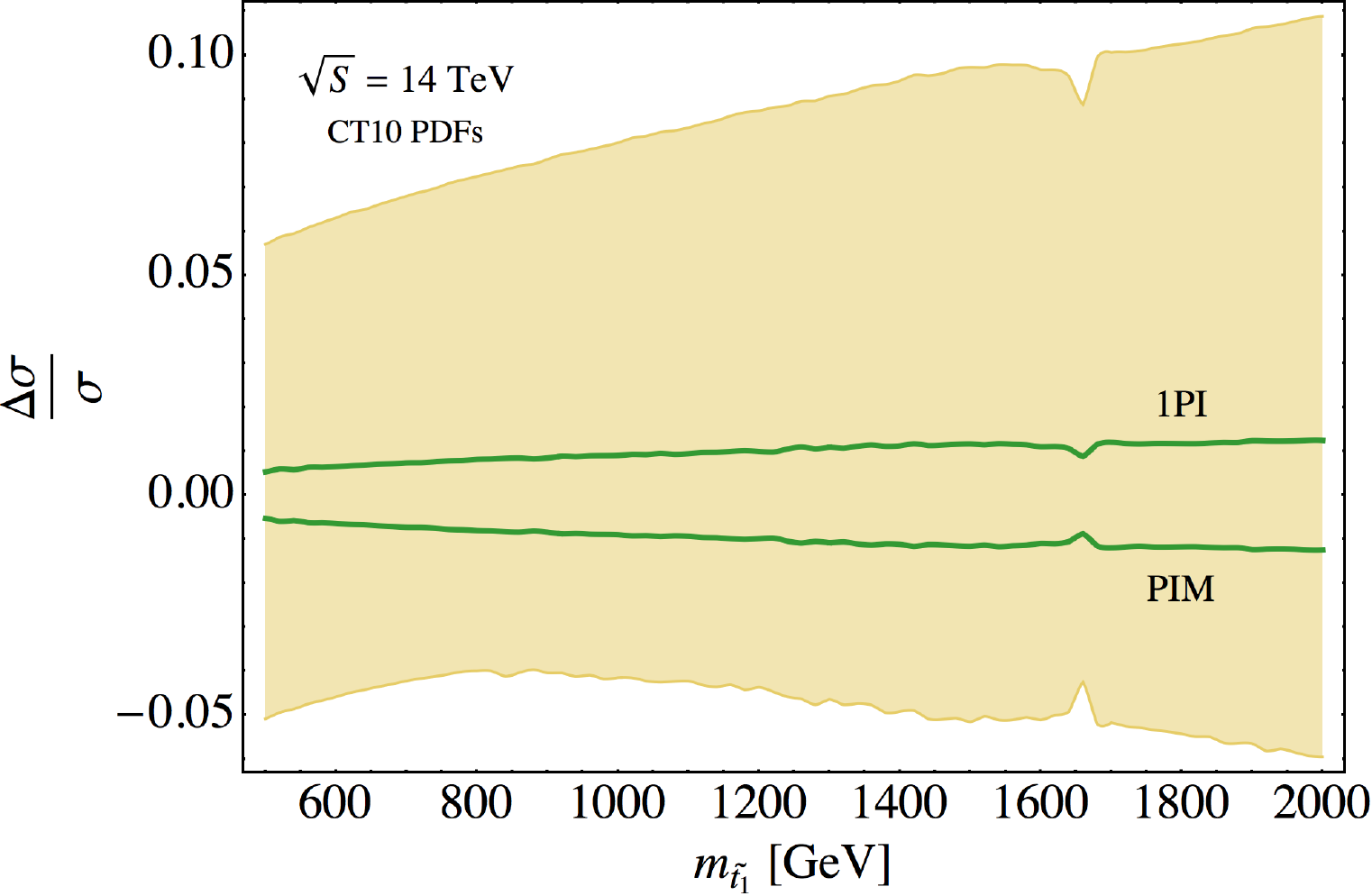} \\ 
\end{tabular} 
\end{center}
\caption{\label{fig:PIMvs1PI} 
Comparison between 1PI and PIM predictions and the residual perturbative uncertainty (brown band) of the averaged prediction (see text for further explanation).}
\end{figure}

Calculations which rely on the use of  PIM and 1PI kinematics neglect different sets of power-suppressed terms and therefore lead to numerically different predictions. In order to account for the scheme uncertainty, we combine the NLO+NNLL predictions in the two kinematic schemes as explained in Section~\ref{sec:Matching}. The differences of the predictions obtained using PIM and 1PI kinematics can be inferred from Figure~\ref{fig:PIMvs1PI}, where the two dark solid lines are obtained by considering, for each value of $m_{\st}$, the quantities
\begin{align}
\frac{\Delta \sigma^{{\tiny \mbox{1PI}}}}{\sigma} =\frac{\sigma^{{\tiny \mbox{1PI}}}_{\tNLO+\tNNLL} - \sigma_{\tNLO+\tNNLL}}{\sigma_{\tNLO+\tNNLL}} & \, \qquad \text{and}
\qquad
\frac{\Delta \sigma^{{\tiny \mbox{PIM}}}}{\sigma} =\frac{\sigma^{{\tiny \mbox{PIM}}}_{\tNLO+\tNNLL} - \sigma_{\tNLO+\tNNLL}}{\sigma_{\tNLO+\tNNLL}} ,
\end{align}
where $\sigma$ without superscript indicates the average between the 1PI and PIM predictions, obtained according to Eq.~(\ref{eq:average}). To obtain these lines all scales (soft, hard, and factorization) are set at their default values discussed in Section~\ref{sec:Matching}. In both panels, the 1PI prediction for the resummed cross section is slightly larger than the PIM prediction in the entire range of values for $m_{\st}$ considered in the figure. However, in both cases the spread between the 1PI and PIM predictions is significantly smaller than the perturbative uncertainty of the combined result, represented by the light brown band and determined as discussed in Section~\ref{sec:Matching}. The slight dent in the bands at $m_{\st}\approx 1660$\,GeV, which is particularly evident in the right panel of the figure, coincides with the gluino--top-quark production threshold.

\subsection{Scale dependence of the resummed cross section}

\begin{figure}[tp]
\begin{center}
\begin{tabular}{cc}
\includegraphics[width=0.48\textwidth]{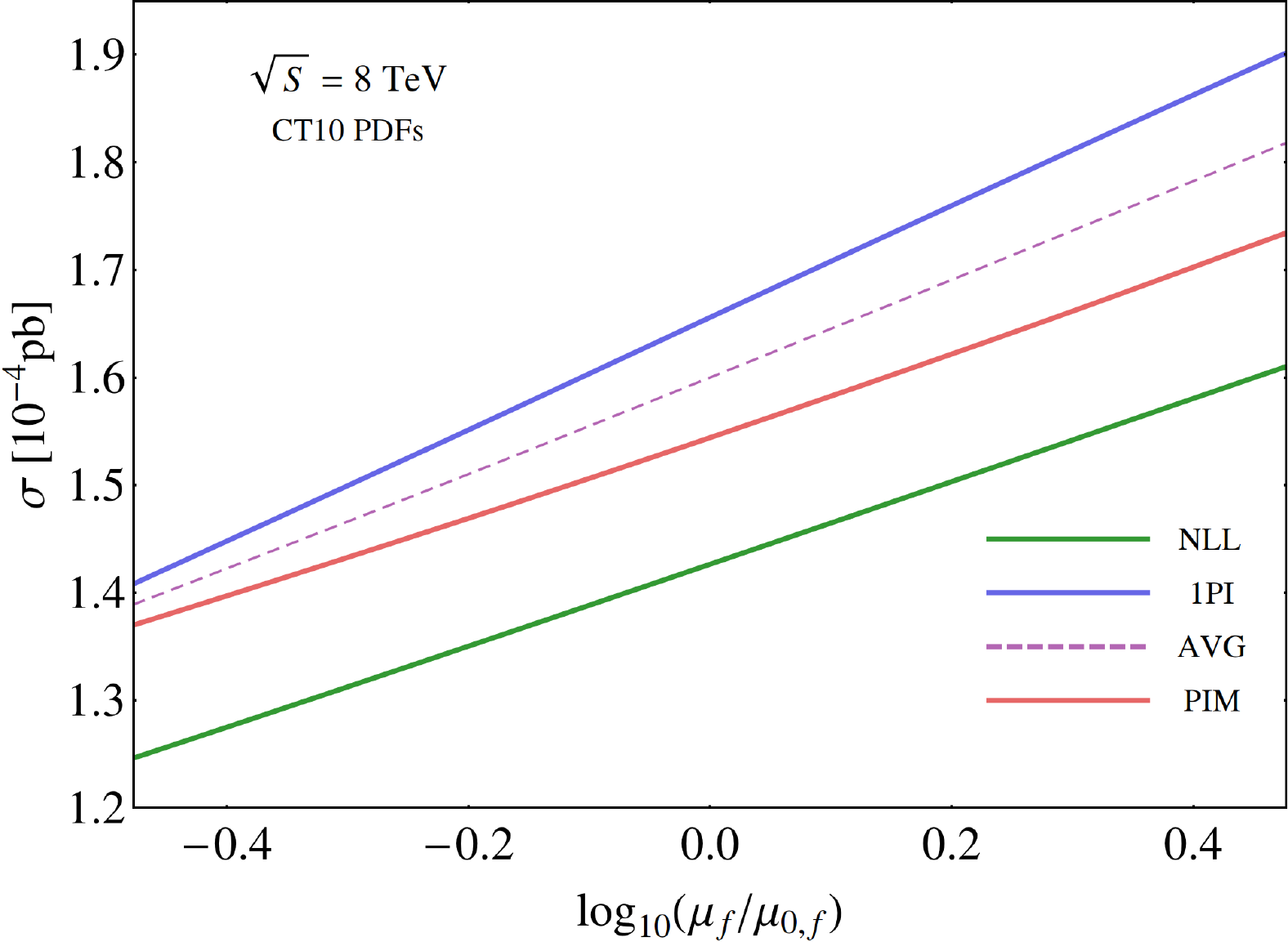} &
\includegraphics[width=0.48\textwidth]{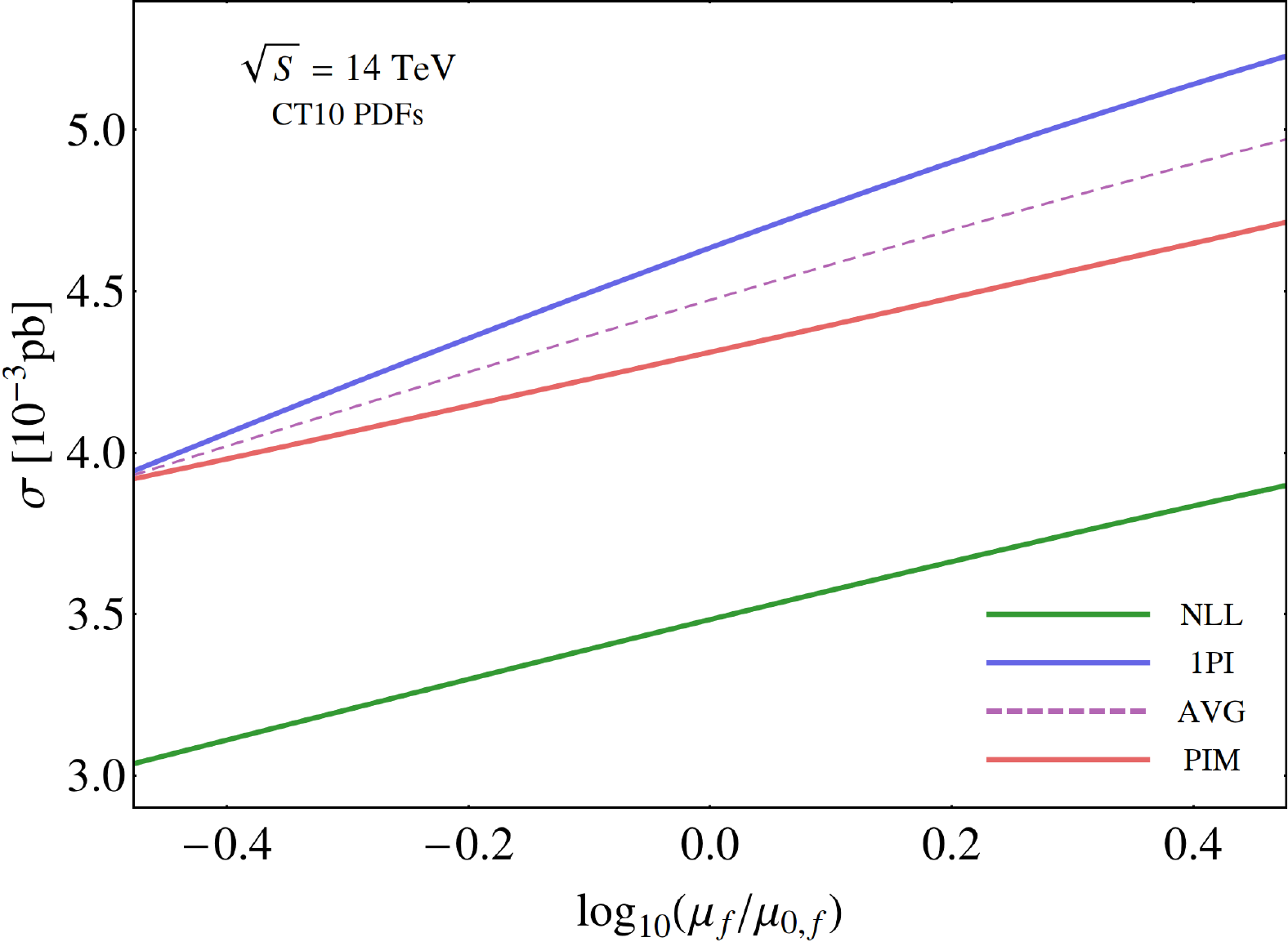} \\
\includegraphics[width=0.48\textwidth]{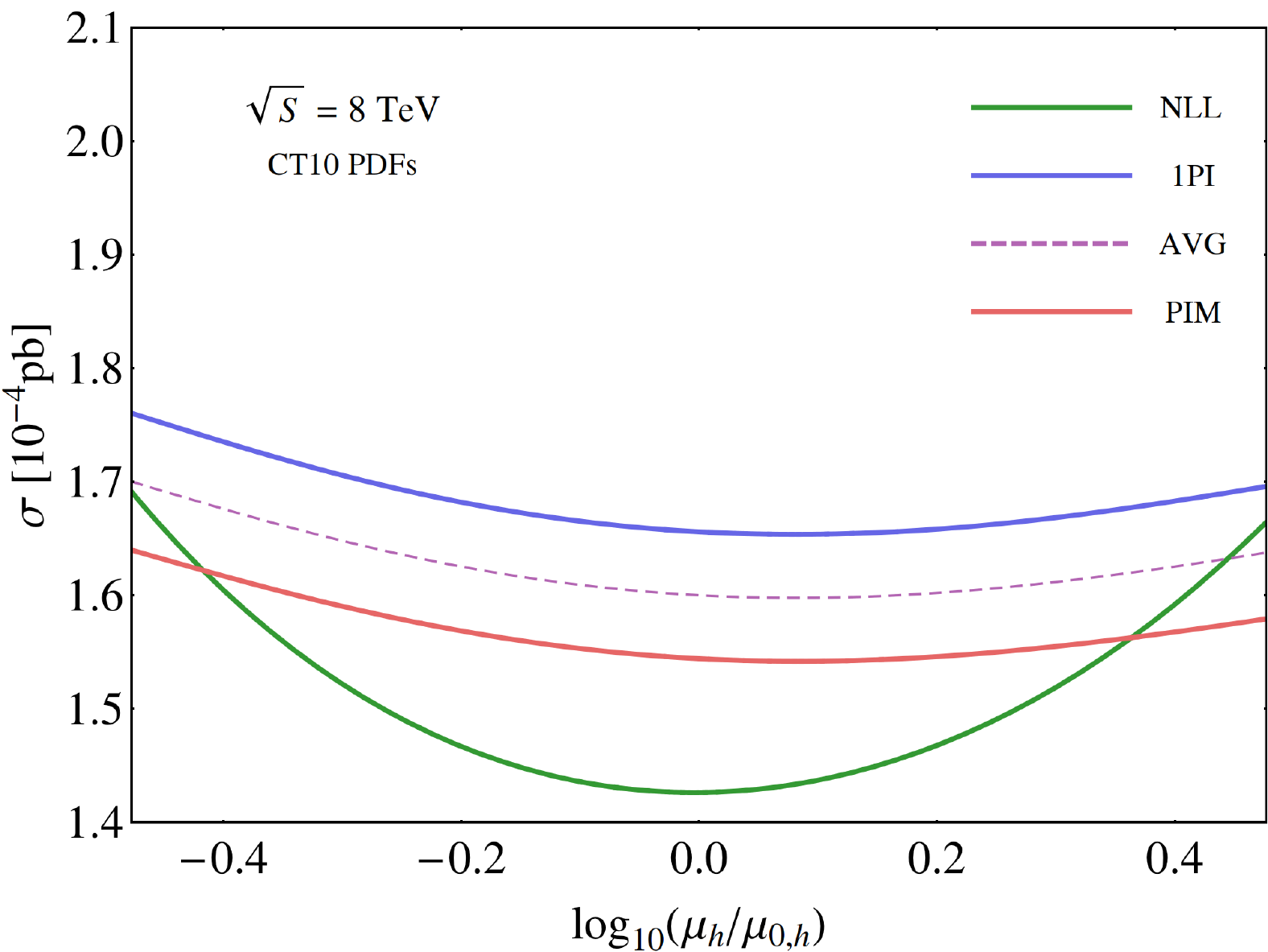} &
\includegraphics[width=0.48\textwidth]{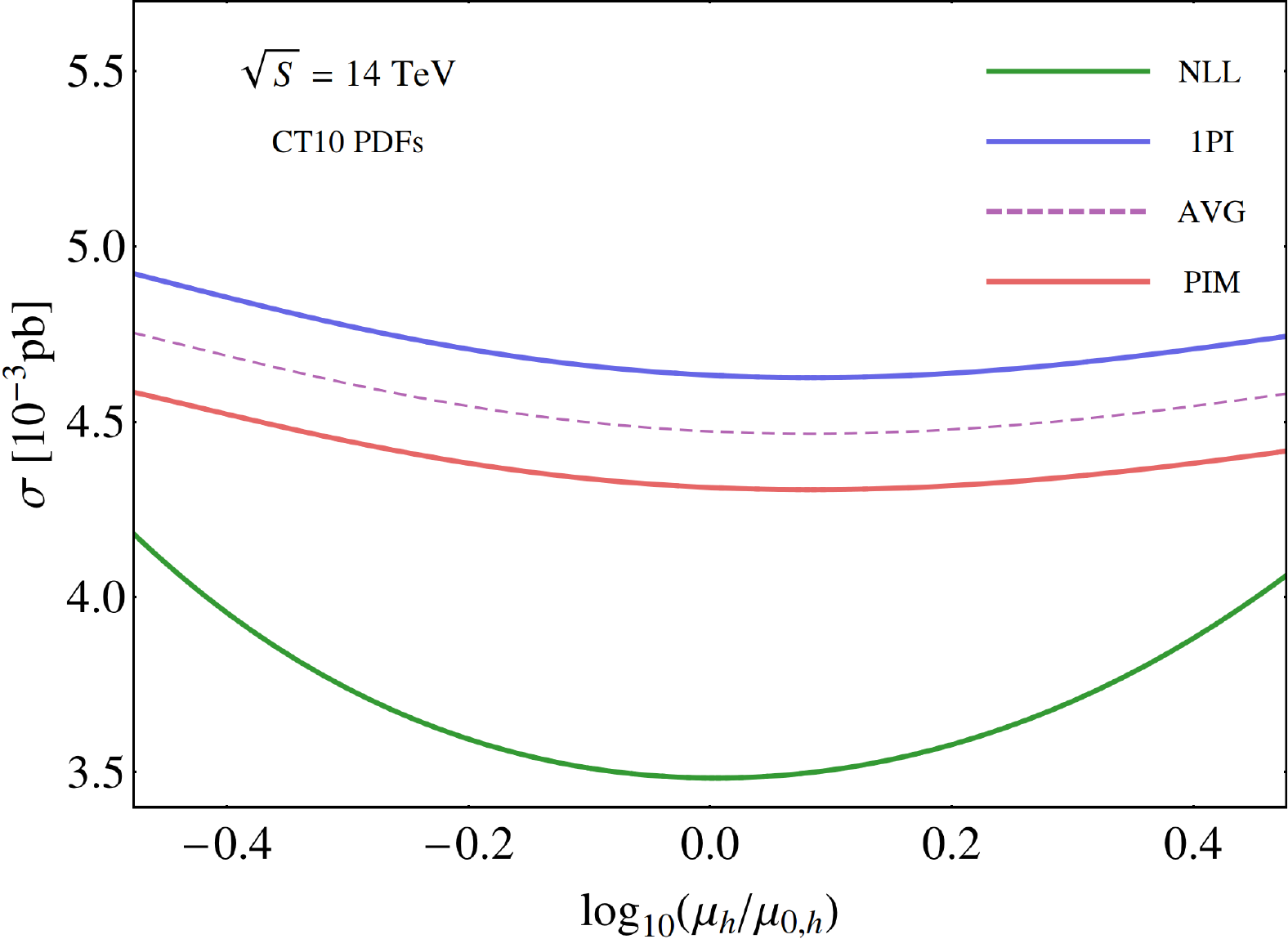} \\
\includegraphics[width=0.48\textwidth]{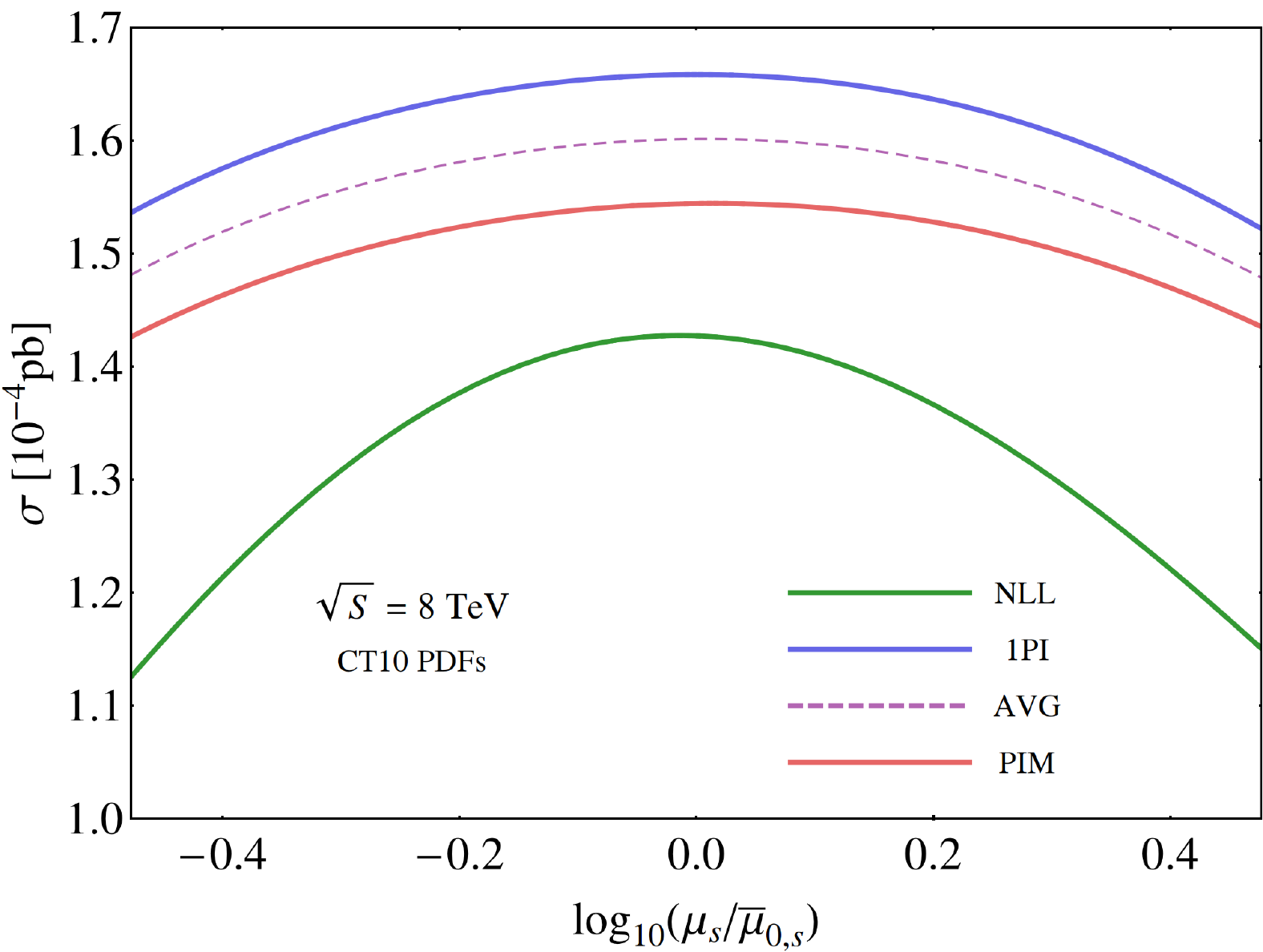} &
\includegraphics[width=0.48\textwidth]{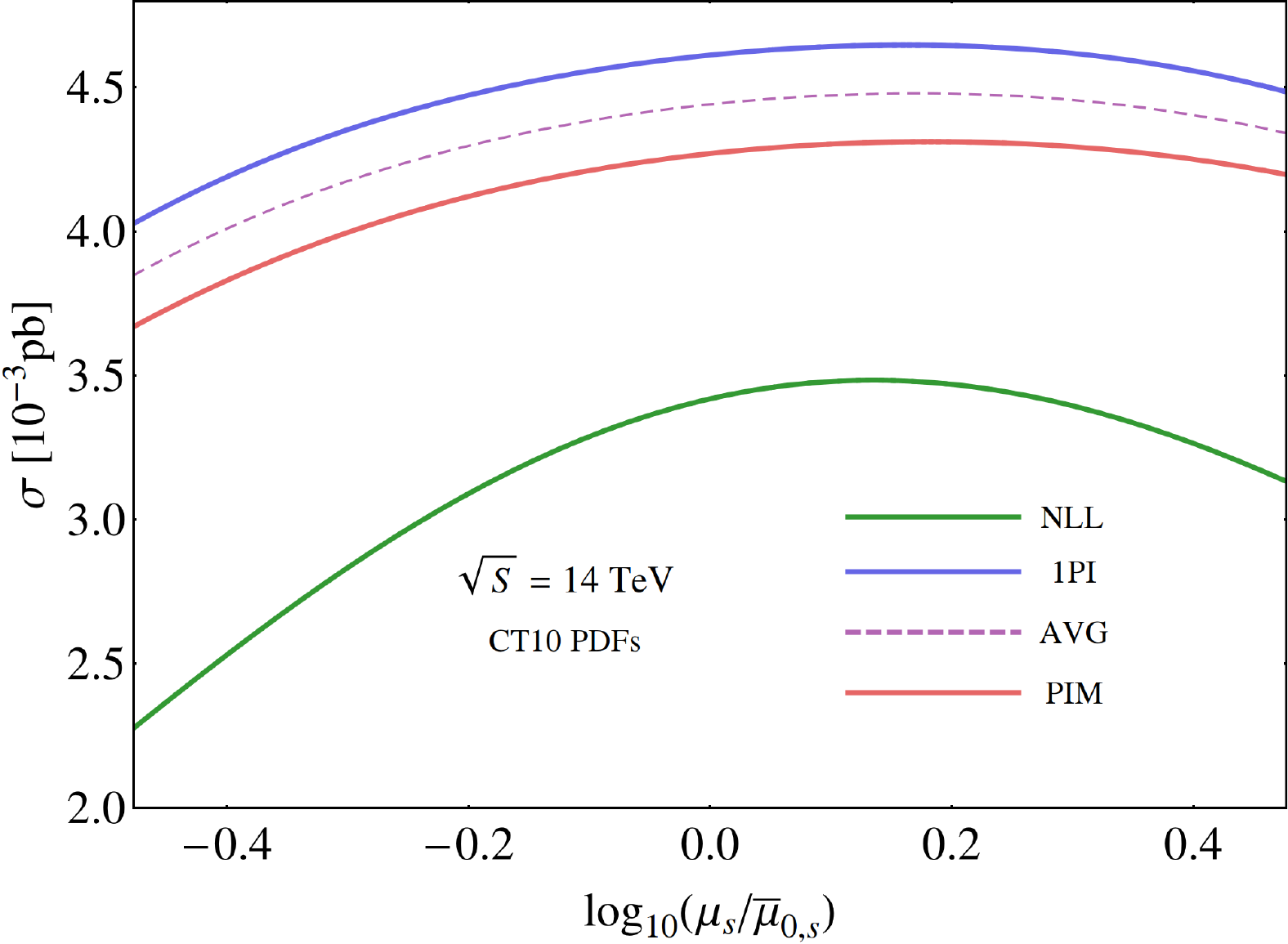} 
\end{tabular} 
\end{center}
\vspace{-0.5cm}
\caption{\label{fig:scalesCT10}
Dependence of the cross section on the factorization scale (first row), hard matching scale (second row), and soft matching scale (third row). The plots in the left panels refer to the LHC at $\sqrt{S}=8$\,TeV, while the right panels refer to 14\,TeV. The reference scales for the factorization and hard scales are chosen equal to their default values $\mu_{0,f}$ and $\mu_{0,h}$. The reference soft scale, $\bar{\mu}_{0,s}$ is set to $250$~GeV  for both collider energies and both kinematics. The three scales are varied in the range $[1/3 \mu_{0,i},3 \mu_{0,i}]$. In order to study the scale dependence of the cross section, the NLL corrections are evaluated using CT10NLO PDFs while the NNLL (non-matched to NLO) corrections are evaluated using CT10NNLO PDFs.}
\end{figure}

An anticipated effect of  the resummation at NNLL order is that phenomenological predictions should be less sensitive to the choice of the soft, hard, and factorization scales when compared to calculations at NLL accuracy. We study this aspect in Figure~\ref{fig:scalesCT10}. In all panels the top-squark mass is set equal to 1087\,GeV. The plots in the left column refer to a LHC center-of-mass energy of 8\,TeV, while the ones on the right column refer to 14\,TeV. 

The two panels in the first row shows the effect of varying the factorization scale about its standard value $\mu_f = m_{\st}$; the soft and factorization scales are kept at their default values for this setup, which are  $\mu_h = 2 m_{\st} =  2174$\,GeV (both in PIM and 1PI kinematics) and $\mu_s = 254$\,GeV (PIM) or $\mu_s = 213$\,GeV (1PI) at $\sqrt{S} = 8$\,TeV, while  $\mu_s = 382$\,GeV (PIM) or $\mu_s = 294$\,GeV (1PI) when $\sqrt{S} = 14$\,TeV. 
The various lines show the scale variation of the NNLL predictions (not matched to NLO) based on PIM and 1PI kinematics, as well as of their average, as detailed in the legend of each panel. For comparison, we also show the scale dependence of the NLL predictions, obtained from an average of the PIM and 1PI cross sections evaluated at that level of accuracy. In Figure~\ref{fig:scalesCT10}, NLL corrections are evaluated by using CT10NLO PDFs, while  NNLL corrections are evaluated by employing CT10NNLO PDFs. By inspecting the plots in the first row, one can see that the cross section at NNLL accuracy has a dependence on $\mu_f$ which is similar to the one of the NLL total stop-production cross section. One encounters a different situation when studying the dependence of the cross section on the hard scale. This fact is illustrated by the plots in the second row of Figure~\ref{fig:scalesCT10}, where the dependence on $\mu_h$ of the cross section at  NNLL and NLL accuracy is shown. In those plots, $\mu_f$ and $\mu_s$ are kept fixed to their default values. One can notice that the various implementations of the NNLL total cross section are less sensitive to the choice of $\mu_h$ than the corresponding calculations at NLL accuracy. In the last row of Figure~\ref{fig:scalesCT10} we consider the dependence of the stop-production cross section on the choice of the soft scale $\mu_s$. The plots show that the NNLL calculations of the cross section span a smaller range of values with respect to NLL calculations when $\mu_s$ is varied. In these two plots, the hard and factorization scales are set to their default values. Finally, all panels in Figure~\ref{fig:scalesCT10} indicate that the NNLL corrections increase the cross section with respect to NLL calculations.

We remind the reader that in order to asses the total scale uncertainty of the 1PI and PIM predictions both at NLL and at NNLL, we first vary each scale in the range $[\mu_{0,i}/2, 2\mu_{0,i}]$ ($i=f,s,h$) and then add the three uncertainties obtained in this way in quadrature. In view of the behavior shown in the plots, one can expect a slightly smaller scale uncertainty at NNLL than at NLL. At this stage we proceed to discuss the effect of the NNLL corrections on the total stop production cross section. 

\subsection{Total Cross Section}

\begin{figure}[t]
\begin{center}
\begin{tabular}{cc}
\includegraphics[width=0.48\textwidth]{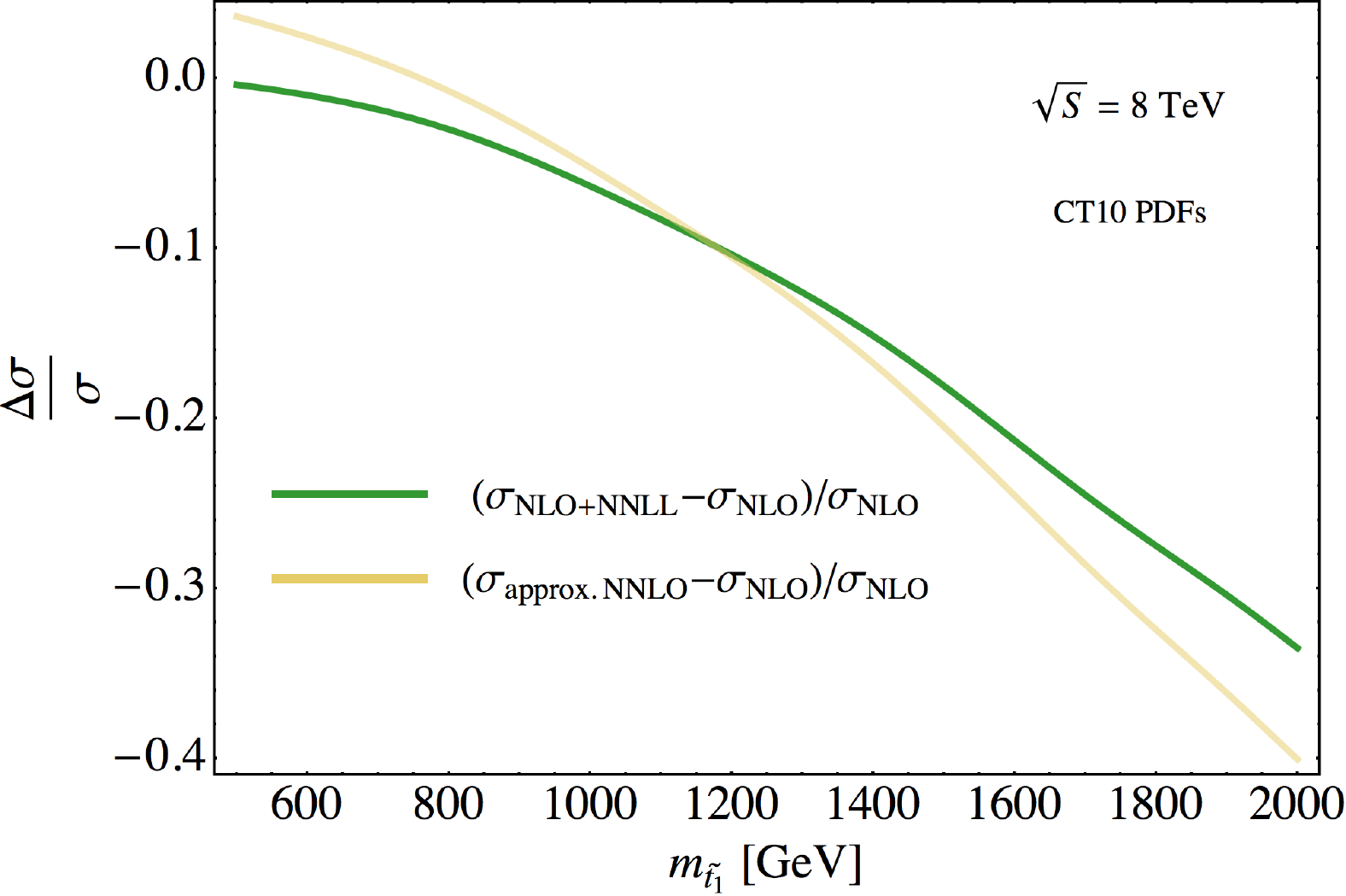} &
\includegraphics[width=0.48\textwidth]{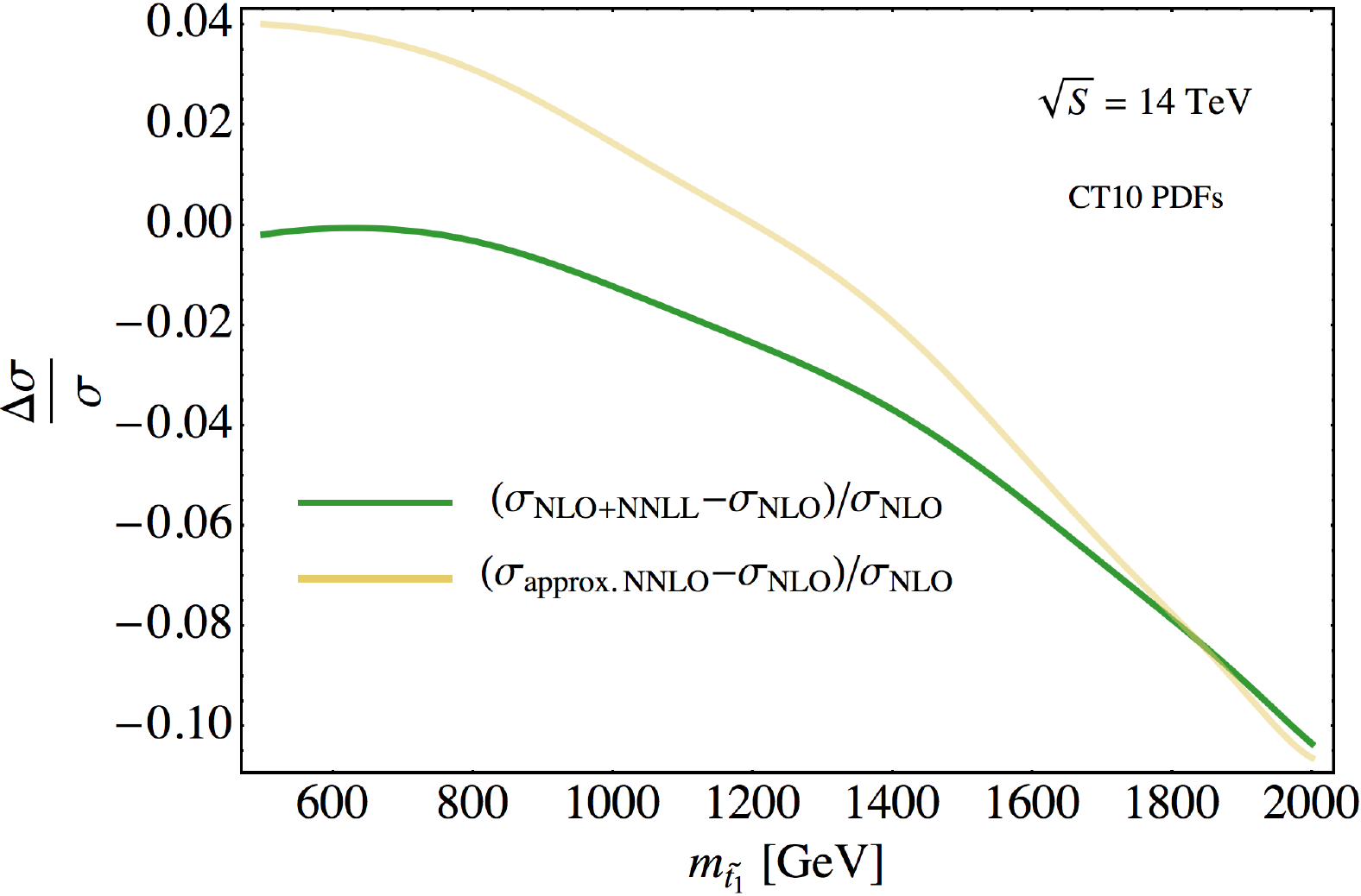}
\end{tabular} 
\end{center}
\caption{\label{fig:NLOvcNNLL8TeVCT10}
Comparison of the relative size of the approximate NNLO and NLO+NNLL corrections with respect to the NLO cross section. The plots span the  mass range $m_{\st} \in [500,2000]$\,GeV. The left and right panels refer to the LHC operating at $\sqrt{S} = 8$\,TeV and $\sqrt{S} = 14$\,TeV, respectively.}
\end{figure}

\begin{table}[th]
\centering
\begin{tabular}{|c||c|c|}
\hline
LHC $8$\,TeV &\multicolumn{2}{|c|}{MSTW2008}  \\
\hline
$m_{\st}$~[GeV] & 500 & 1087.17 \\
\hline
\hline
$(\sigma \pm \Delta \sigma_\mu \pm \Delta_{\rm{PDF} })_{\rm{{\tiny LO}}}$ [pb] & $61.7^{+27.3 + 6.1 }_{-17.5 -6.0}  \times 10^{-3}$  & $11.5^{+5.6 + 2.5 }_{-3.5 -2.0}\times 10^{-5}$\\
\hline
$(\sigma \pm \Delta \sigma_\mu \pm \Delta_{\rm{PDF} + \alpha_s})_{\rm{{\tiny NLO}}}$ [pb] &$83.4^{+10.5 + 10.6}_{-12.2 -8.8} \times 10^{-3}$  & $ 14.7^{+2.1+ 3.7}_{-2.5 -2.8}\times 10^{-5}$ \\
\hline
$(\sigma \pm \Delta \sigma_\mu\pm \Delta_{\rm{PDF} + \alpha_s})_{\rm{{\tiny NLL}}}$ [pb] & $ 62.6^{+12.1+7.2}_{-11.4-6.1} \times 10^{-3}$ & $13.0^{+2.1+3.1}_{-2.0-2.6} \times 10^{-5}$ \\
\hline
$(\sigma \pm \Delta \sigma_\mu\pm \Delta_{\rm{PDF} + \alpha_s})_{\rm{{\tiny approx.\ NNLO}}}$ [pb] & $ 83.2^{+3.3 + 12.6}_{-4.9-9.9} \times 10^{-3}$ & $ 15.3^{+0.3+5.8}_{-1.0-3.0} \times 10^{-5}$ \\
\hline
$(\sigma \pm \Delta \sigma_\mu\pm \Delta_{\rm{PDF} + \alpha_s})_{\rm{{\tiny NLO+NNLL}}}$ [pb] & $79.9^{+6.2 + 12.0}_{-3.9 -9.3} \times 10^{-3}$ & $15.2^{+1.7+5.7}_{-0.9-2.9} \times 10^{-5}$\\
\hline
$K_{\rm{NLO}}$ & $1.35$ & $1.29$  \\
\hline
$K_{\rm{NLL}}$ & $1.01$ & $1.13$  \\
\hline
$K_{\rm{approx.\ NNLO}}$ & $1.35$ & $1.34$ \\
\hline
$K_{\rm{NLO+NNLL}}$ & $1.29$  &  $1.32$ \\
\hline
\end{tabular}
\caption{Stop-pair production cross section for two different values of $m_{\tilde{t}_1}$ at the LHC with $\sqrt{S} = 8$\,TeV. The numbers are obtained by using MSTW2008 PDFs.
\label{tab:8tevMSTW.40.2.5}}
\end{table}

\begin{table}[th]
\centering
\begin{tabular}{|c||c|c|}
\hline
LHC $8$\,TeV &\multicolumn{2}{|c|}{CT10}  \\
\hline
$m_{\st}$~[GeV] & 500 & 1087.17 \\
\hline
\hline
$(\sigma \pm \Delta \sigma_\mu \pm \Delta_{\rm{PDF} + \alpha_s})_{\rm{{\tiny LO}}}$ [pb] & $54.0^{+21.2 + 11.0}_{-14.2 -8.3} \times 10^{-3}$ & $10.6^{+4.8 +6.6}_{-3.1-3.2}\times 10^{-5}$ \\
\hline
$(\sigma \pm \Delta \sigma_\mu \pm \Delta_{\rm{PDF} + \alpha_s})_{\rm{{\tiny NLO}}}$ [pb] & $80.9^{+9.8 + 16.6}_{-11.4 -13.1} \times 10^{-3}$  & $ 16.5^{+2.3+ 10.4}_{-2.7-5.3} \times 10^{-5}$ \\
\hline
$(\sigma \pm \Delta \sigma_\mu\pm \Delta_{\rm{PDF} + \alpha_s})_{\rm{{\tiny NLL}}}$ [pb] & $ 60.6^{+11.7+11.4}_{-10.7-9.0} \times 10^{-3}$ & $ 14.3^{+2.3+8.1}_{-2.2-4.2}\times 10^{-5}$ \\
\hline
$(\sigma \pm \Delta \sigma_\mu\pm \Delta_{\rm{PDF} + \alpha_s})_{\rm{{\tiny approx.\ NNLO}}}$ [pb] & $ 83.6^{+3.6+ 19.0}_{-4.8-12.3} \times 10^{-3}$ &  $ 15.2^{+0.3+ 8.1}_{-1.0-4.7} \times 10^{-5}$ \\
\hline
$(\sigma \pm \Delta \sigma_\mu\pm \Delta_{\rm{PDF} + \alpha_s})_{\rm{{\tiny NLO+NNLL}}}$ [pb] & $80.5^{+6.3+17.5}_{-4.0-12.2} \times 10^{-3}$ & $15.1^{+1.8+8.0}_{-1.0-4.6} \times 10^{-5}$\\
\hline
$K_{\rm{NLO}}$ &$1.50$  & $1.56$  \\
\hline
$K_{\rm{NLL}}$ &$1.12$  & $1.35$  \\
\hline
$K_{\rm{approx.\ NNLO}}$ &$1.55$  & $1.44$  \\
\hline
$K_{\rm{NLO+NNLL}}$ & $1.49$ & $1.42$  \\
\hline
\end{tabular}
\caption{Stop-pair production cross section for two different values of $m_{\tilde{t}_1}$ at the LHC with $\sqrt{S} = 8$\,TeV. The numbers are obtained by using CT10 PDFs.
\label{tab:8tevCT10.40.2.5}}
\end{table}

\begin{table}[th]
\centering
\begin{tabular}{|c||c|c|}
\hline
LHC $14$\,TeV &\multicolumn{2}{|c|}{MSTW2008} \\
\hline
$m_{\st}$~[GeV] & 500 & 1087.17 \\
\hline
\hline
$(\sigma \pm \Delta \sigma_\mu \pm \Delta_{\rm{PDF} })_{\rm{{\tiny LO}}}$ [pb] & $48.3^{+18.4+3.3}_{-12.4-3.4} \times 10^{-2}$ & $ 33.5^{+13.8+3.7}_{-9.1-3.6} \times 10^{-4}$\\
\hline
$(\sigma \pm \Delta \sigma_\mu \pm \Delta_{\rm{PDF} + \alpha_s})_{\rm{{\tiny NLO}}}$ [pb] & $66.4^{+7.7+6.2}_{-8.5-5.2} \times 10^{-2}$ & $ 44.2^{+4.9+6.4}_{-6.0-5.1} \times 10^{-4}$ \\
\hline
$(\sigma \pm \Delta \sigma_\mu \pm \Delta_{\rm{PDF} + \alpha_s})_{\rm{{\tiny NLL}}}$ [pb] & $46.9^{+9.8+3.8}_{-8.7-3.2}\times 10^{-2}$ & $ 35.1^{+6.2+4.8}_{-5.6-3.8}\times 10^{-4}$ \\
\hline
$(\sigma \pm \Delta \sigma_\mu\pm \Delta_{\rm{PDF} + \alpha_s})_{\rm{{\tiny approx.\ NNLO}}}$ [pb] & $65.7^{+3.3+6.5}_{-3.4-6.2}\times 10^{-2}$ & $ 44.3^{+1.3+7.8}_{-2.2-5.4} \times 10^{-4}$  \\
\hline
$(\sigma \pm \Delta \sigma_\mu\pm \Delta_{\rm{PDF} + \alpha_s})_{\rm{{\tiny NLO+NNLL}}}$ [pb] & $62.9^{+3.5+6.2}_{-3.2-5.6} \times 10^{-2}$  & $43.1^{+3.5+7.4}_{-1.8-5.1} \times 10^{-4}$ \\
\hline
$K_{\rm{NLO}}$ & $1.38$  & $1.32$  \\
\hline
$K_{\rm{NLL}}$ & $0.97$  & $1.05$  \\
\hline
$K_{\rm{approx.\ NNLO}}$ & $1.36$ &  $1.32$\\
\hline
$K_{\rm{NLO+NNLL}}$ & $1.30$ & $1.29$  \\
\hline
\end{tabular}
\caption{Stop-pair production cross section for two different values of $m_{\tilde{t}_1}$ at the LHC with $\sqrt{S} = 14$\,TeV. The numbers are obtained by using MSTW2008 PDFs.
\label{tab:14tevMSTW.40.2.5}}
\end{table}

\begin{table}[th]
\centering
\begin{tabular}{|c||c|c||}
\hline
LHC $14$\,TeV &\multicolumn{2}{|c|}{CT10}  \\
\hline
$m_{\st}$~[GeV] & 500 & 1087.17 \\
\hline
\hline
$(\sigma \pm \Delta \sigma_\mu \pm \Delta_{\rm{PDF} + \alpha_s})_{\rm{{\tiny LO}}}$ [pb] & $42.6^{+14.4+5.0}_{-10.1-4.3} \times 10^{-2}$ & $ 30.1^{+11.3+ 7.8}_{-7.7 -5.2} \times 10^{-4}$ \\
\hline
$(\sigma \pm \Delta \sigma_\mu \pm \Delta_{\rm{PDF} + \alpha_s})_{\rm{{\tiny NLO}}}$ [pb] & $63.2^{+7.0+7.6}_{-7.8-6.6} \times 10^{-2}$ & $ 44.1^{+4.8+11.7}_{-5.8-8.1} \times 10^{-4}$ \\
\hline
$(\sigma \pm \Delta \sigma_\mu \pm \Delta_{\rm{PDF} + \alpha_s})_{\rm{{\tiny NLL}}}$ [pb] & $44.7^{+9.3+4.9}_{-8.2-4.2}\times 10^{-2}$ & $ 34.8^{+6.1+8.4}_{-5.5-5.9} \times 10^{-4}$ \\
\hline
$(\sigma \pm \Delta \sigma_\mu\pm \Delta_{\rm{PDF} + \alpha_s})_{\rm{{\tiny approx.\ NNLO}}}$ [pb] & $65.9^{+3.4+8.2}_{-3.4-6.6} \times 10^{-2}$ & $ 44.6^{+1.3 + 12.1}_{-2.1-7.8} \times 10^{-4}$  \\
\hline
$(\sigma \pm \Delta \sigma_\mu\pm \Delta_{\rm{PDF} + \alpha_s})_{\rm{{\tiny NLO+NNLL}}}$ [pb] & $63.1^{+3.5 +7.7}_{-3.4-6.3} \times 10^{-2}$ & $43.4^{+3.6+11.5}_{-1.8-7.6} \times 10^{-4}$ \\
\hline
$K_{\rm{NLO}}$ & $1.48$ & $1.47$  \\
\hline
$K_{\rm{NLL}}$ & $1.05$ & $1.16$  \\
\hline
$K_{\rm{approx.\ NNLO}}$ & $1.55$ & $1.48$  \\
\hline
$K_{\rm{NLO+NNLL}}$ &  $1.48$ & $1.44$   \\
\hline
\end{tabular}
\caption{Stop-pair production cross section for two different values of $m_{\tilde{t}_1}$ at the LHC with  $\sqrt{S} = 14$\,TeV. The numbers are obtained by using CT10 PDFs.
\label{tab:14tevCT10.40.2.5}}
\end{table}

We now proceed to study the effect of the resummation at NNLL accuracy on the total stop pair production cross section. We study this aspect by comparing the NNLL predictions for the cross section (matched to the fixed-order NLO cross section) with NLO, NLL, and approximate NNLO predictions for the same observable. In this section, LO predictions are obtained by employing LO PDFs, NLO and NLL calculations employ NLO PDFs, while approximate NNLO and NLO+NNLL calculations are carried out by using NNLO PDFs. We remind the reader that PDFs at different orders (and, consequently, the cross-section predictions) employ different values of $\alpha_s$. 

We start by discussing the stop production cross section at two different values of the top-squark mass: $m_{\st} = 500$\,GeV and $m_{\st} = 1087$\,GeV. Tables~\ref{tab:8tevMSTW.40.2.5} and \ref{tab:8tevCT10.40.2.5} show the predictions for the LHC operating at a center-of-mass energy of 8\,TeV, obtained using the PDF sets MSTW2008 and CT10, respectively. Tables~\ref{tab:14tevMSTW.40.2.5} and \ref{tab:14tevCT10.40.2.5} show the corresponding results for the case in which $\sqrt{S} = 14$\,TeV. The $K$-factors are defined as
\begin{align}
K_i = \frac{\sigma_i}{\sigma_{\tiny\rm LO}} \,; \qquad 
\mbox{with $i=$NLO, NLL, approx.\ NNLO, NLO+NNLL.}
\end{align}
Inspecting the tables, we observe that the NLL calculations predict a smaller cross section than the NLO calculations. This effect is particularly pronounced at $m_{\st} = 500$\,GeV. This means that NLO contributions not included in NLL soft gluon emission corrections are numerically sizable, in particular for smaller values of the stop mass, where hard gluon emission is less suppressed by phase-space constraints. A similar behavior was already encountered in the study of the top-quark pair production cross section (see for example Table~4 in \cite{Ahrens:2010zv}).

\begin{figure}[t]
\begin{center}
\begin{tabular}{cc}
  \includegraphics[width=0.48\textwidth]{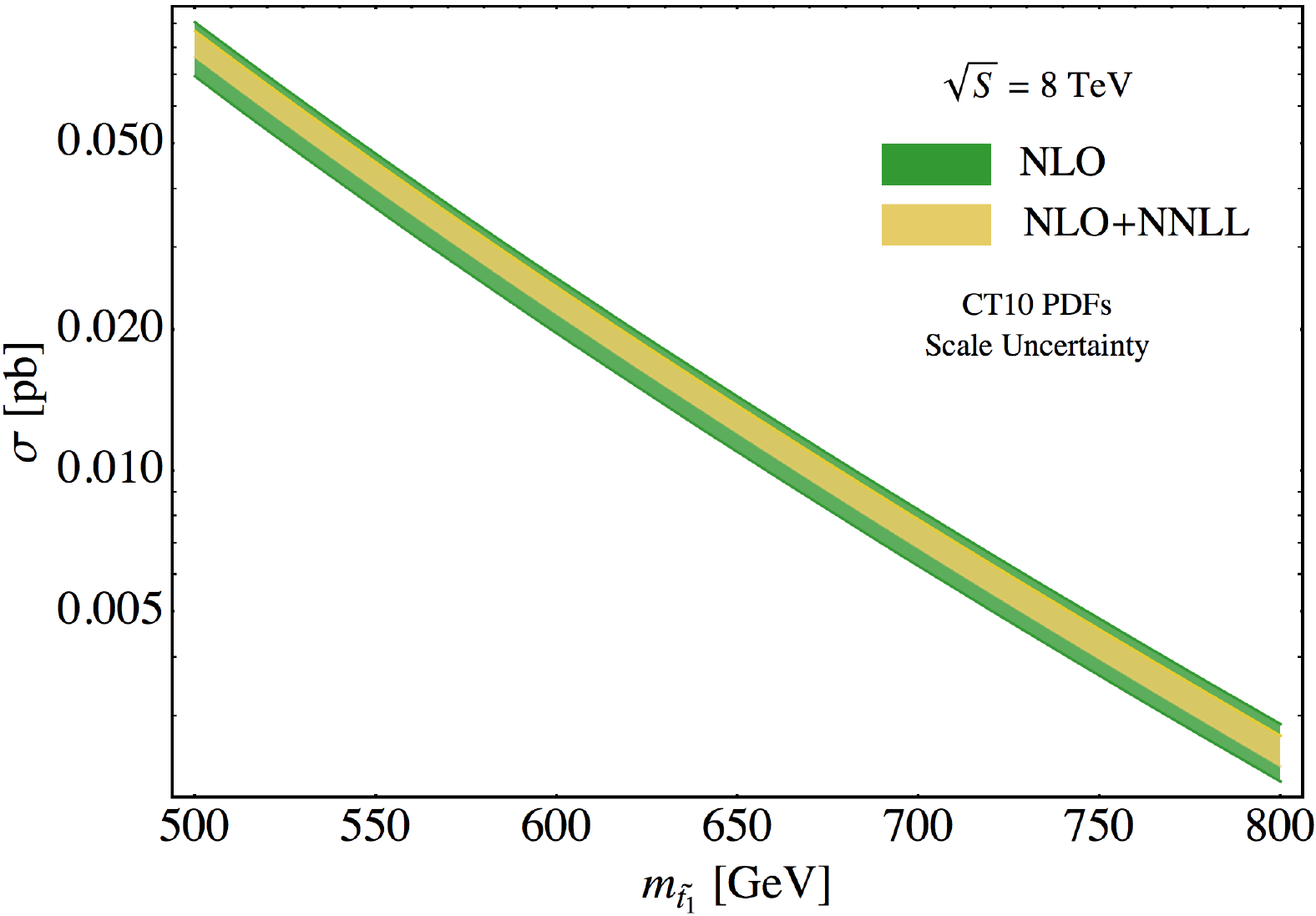} & \includegraphics[width=0.48\textwidth]{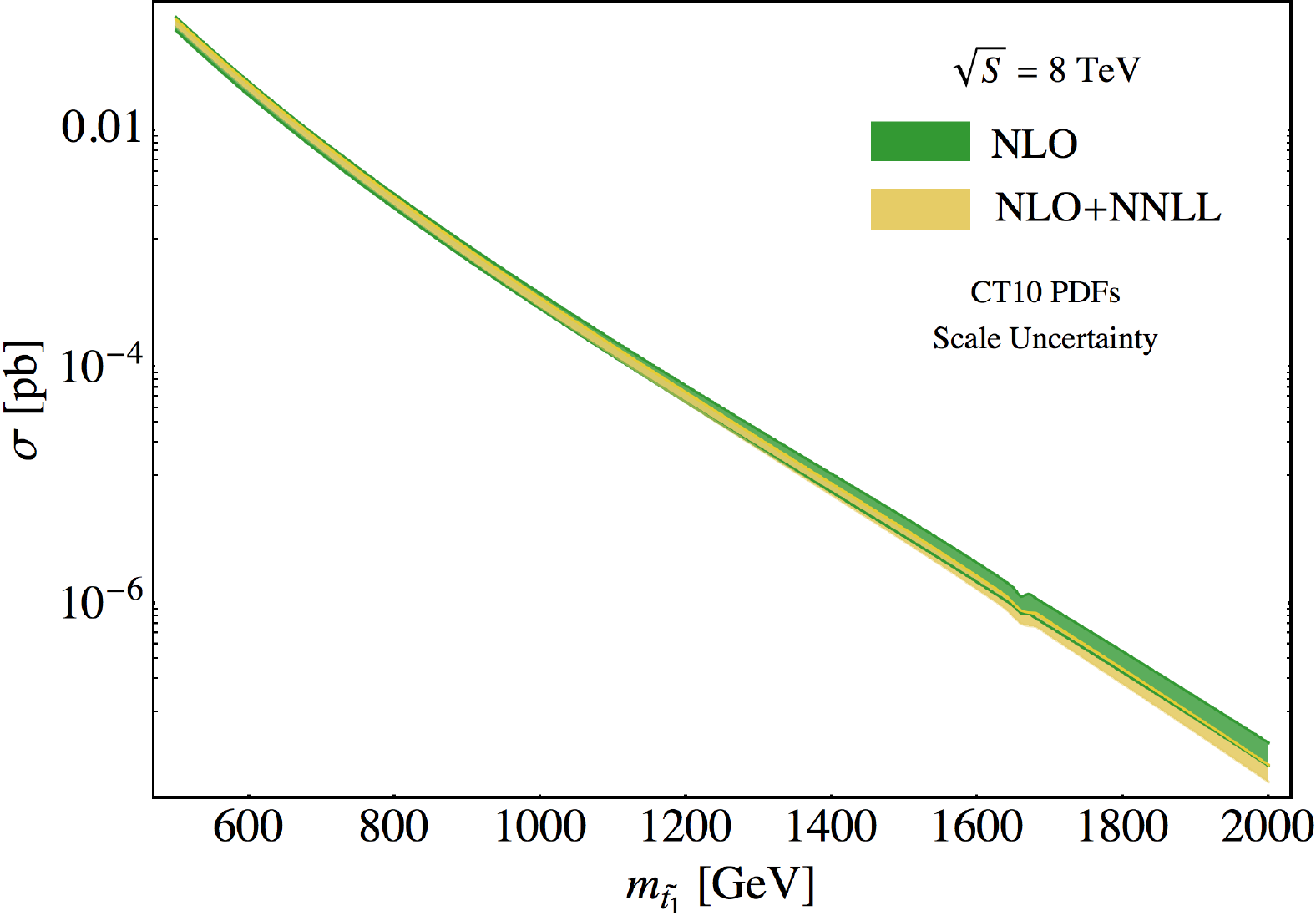} \\
 \includegraphics[width=0.48\textwidth]{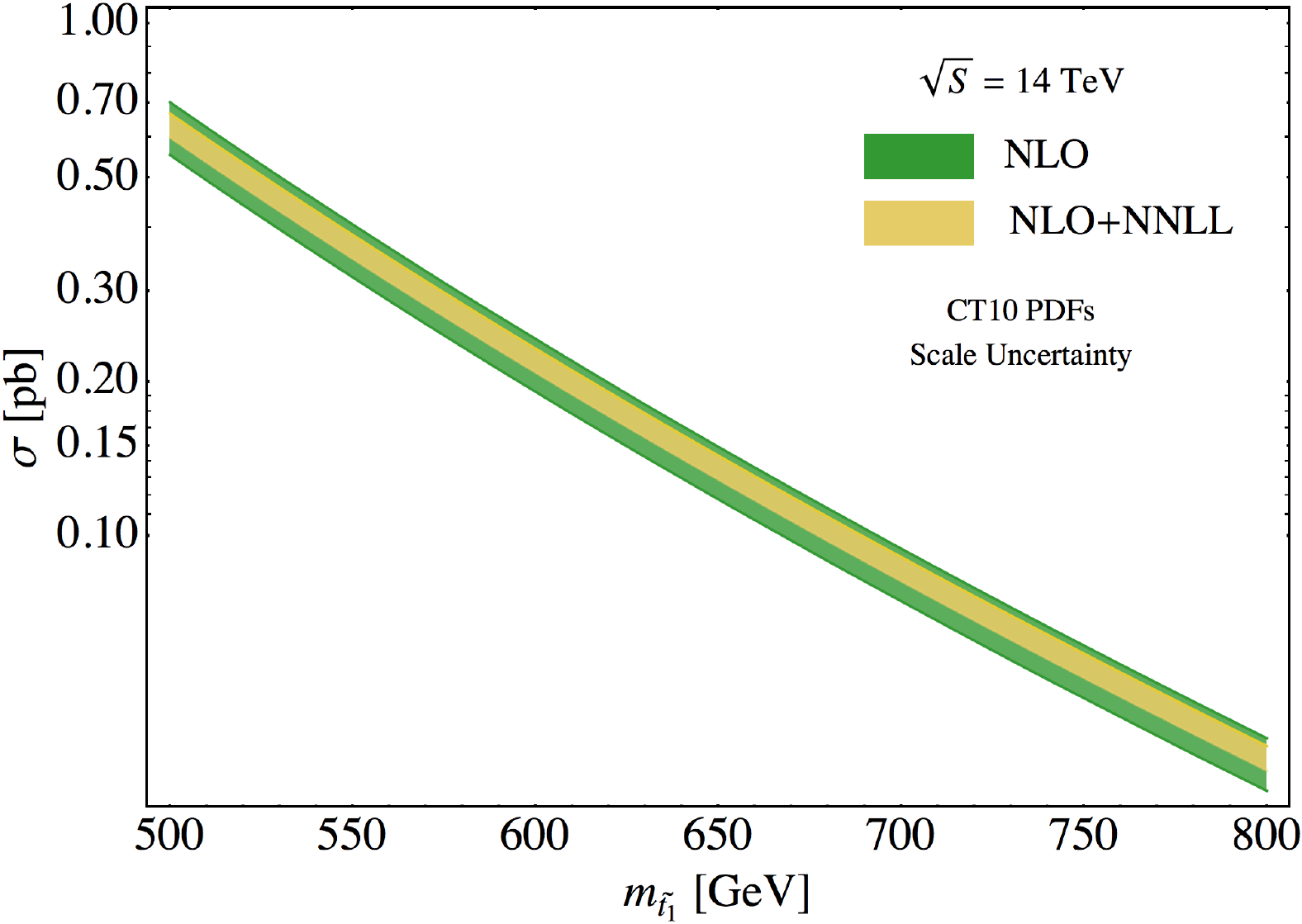}  & \includegraphics[width=0.49\textwidth]{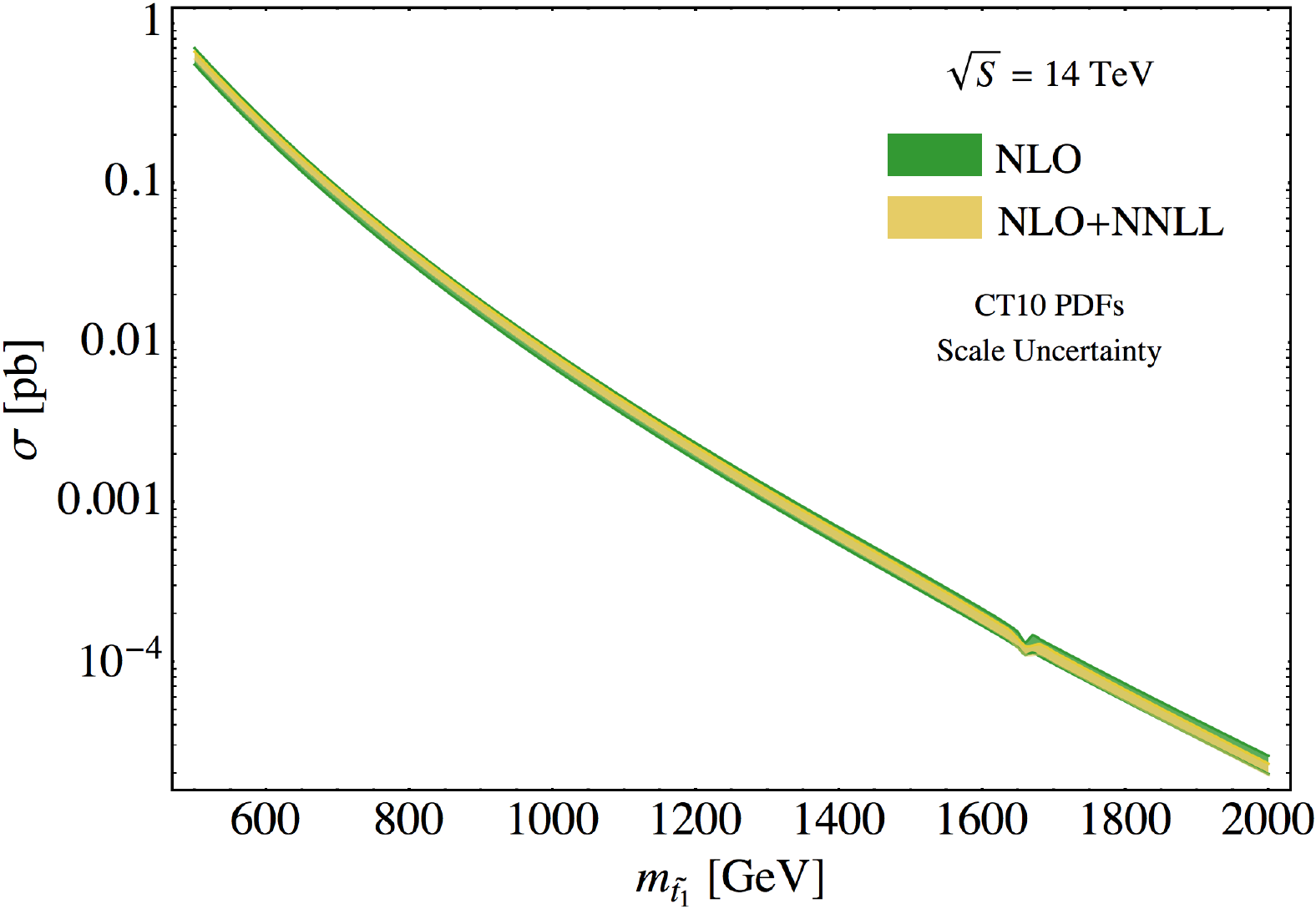} 
\end{tabular} 
\end{center}
\caption{\label{fig:massscan8TCT10} 
Mass scans with CT10 PDFs for the LHC with $\sqrt{S}=8$\,TeV (first row) and $\sqrt{S}=14$\,TeV (second row). The bands represent the perturbative scale uncertainties at NLO and NLO+NNLL. The left panels show a detail of the mass range 500\textemdash800~GeV. All of the SUSY parameters other than $m_{\tilde{t}_1}$ are fixed at the values of the benchmark point {\tt 40.2.5} \cite{AbdusSalam:2011fc}. The plots are obtained by employing CT10NNLO PDFs \cite{Lai:2010vv,Gao:2013xoa}.}
\end{figure}

\begin{figure}[t]
\begin{center}
\begin{tabular}{cc}
  \includegraphics[width=0.48\textwidth]{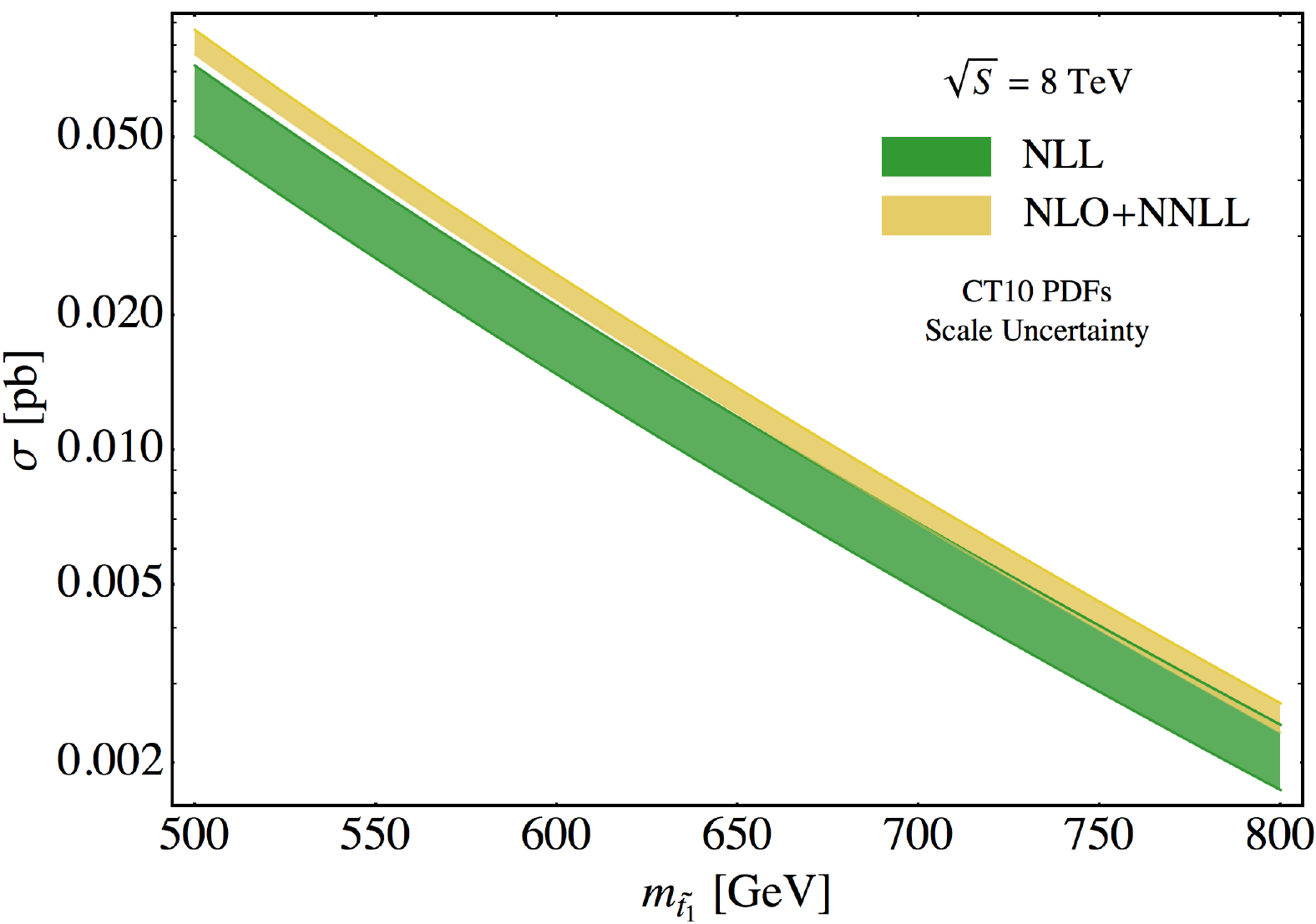} & \includegraphics[width=0.48\textwidth]{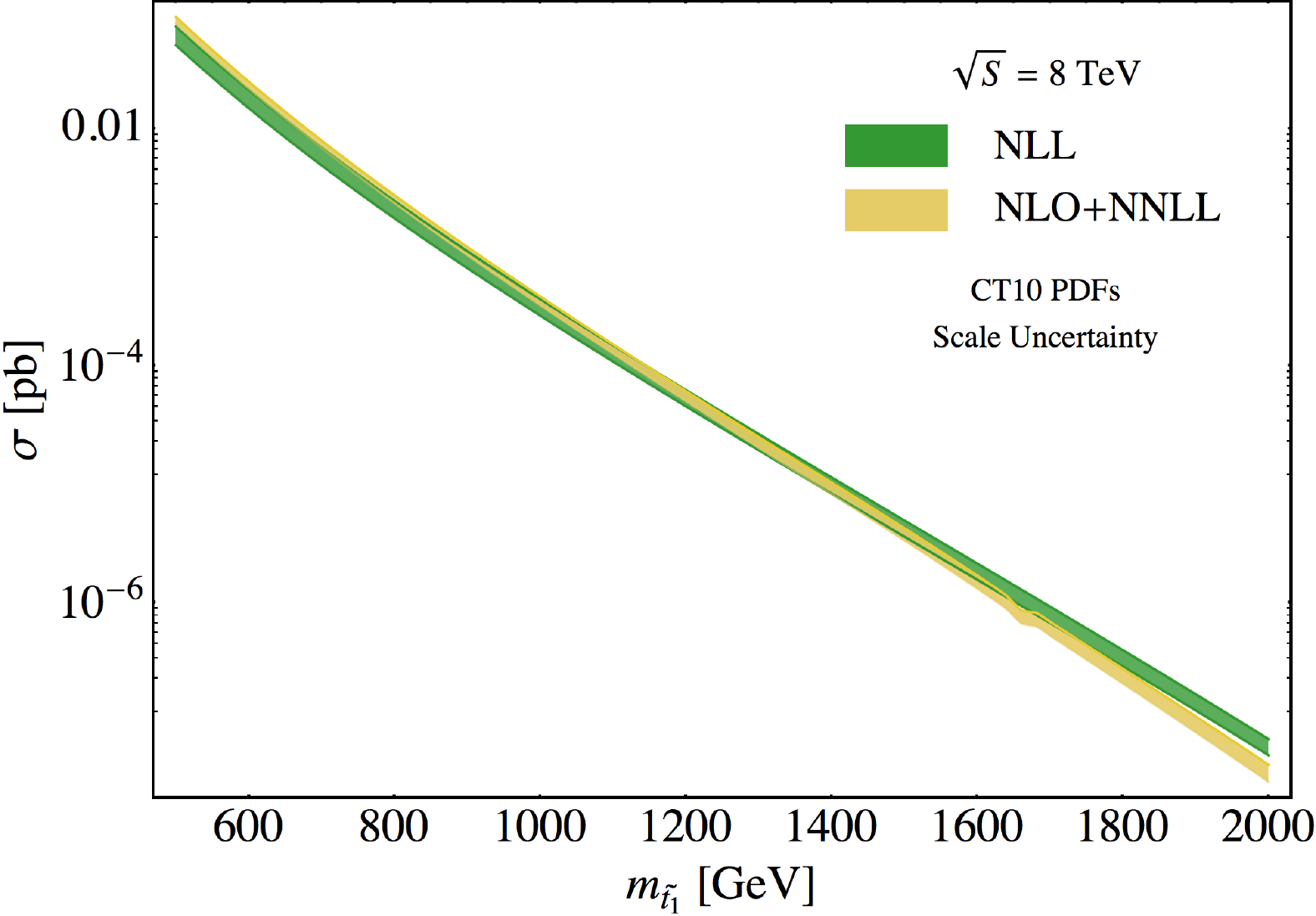} \\
\includegraphics[width=0.48\textwidth]{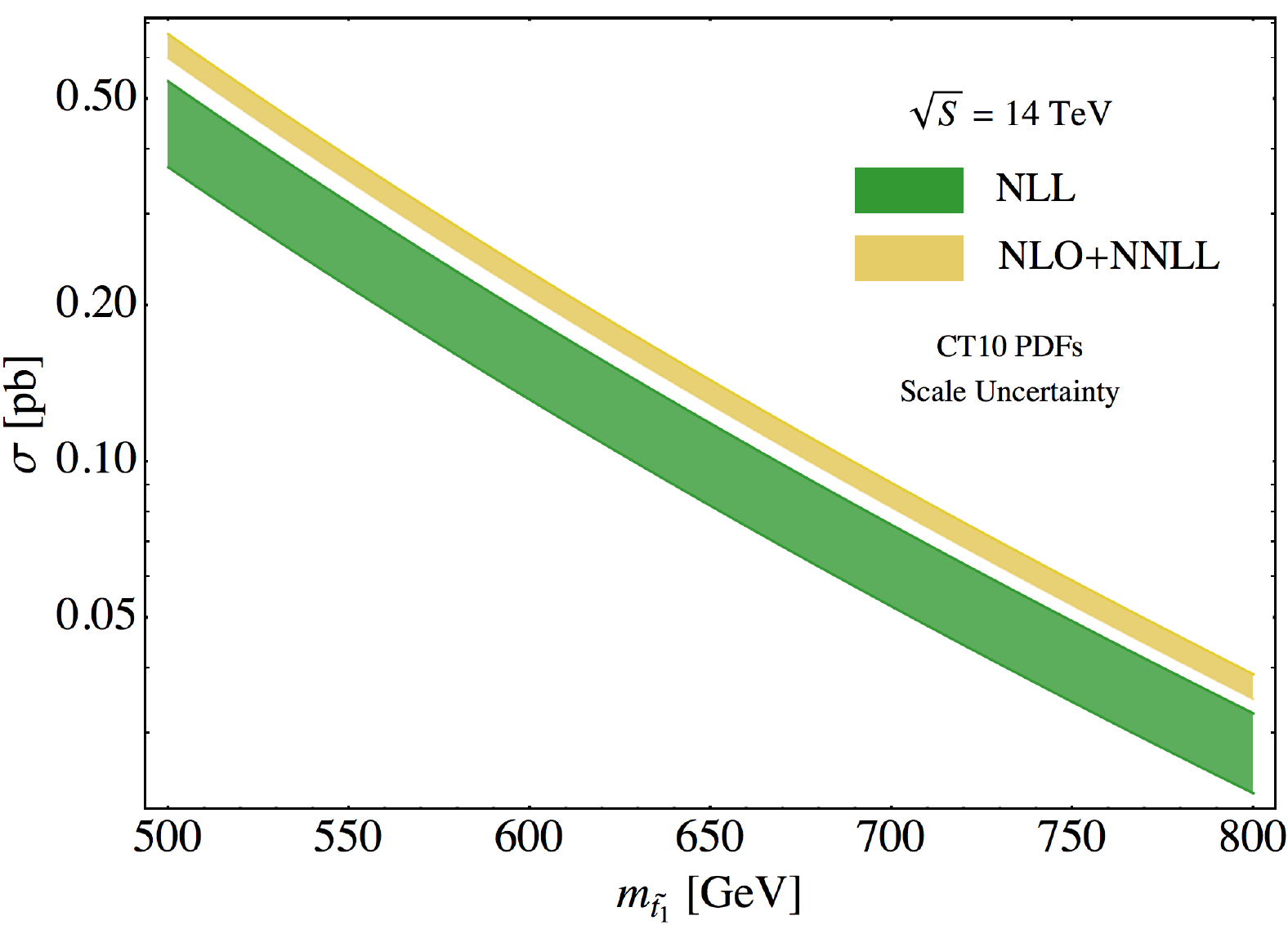}  & \includegraphics[width=0.49\textwidth]{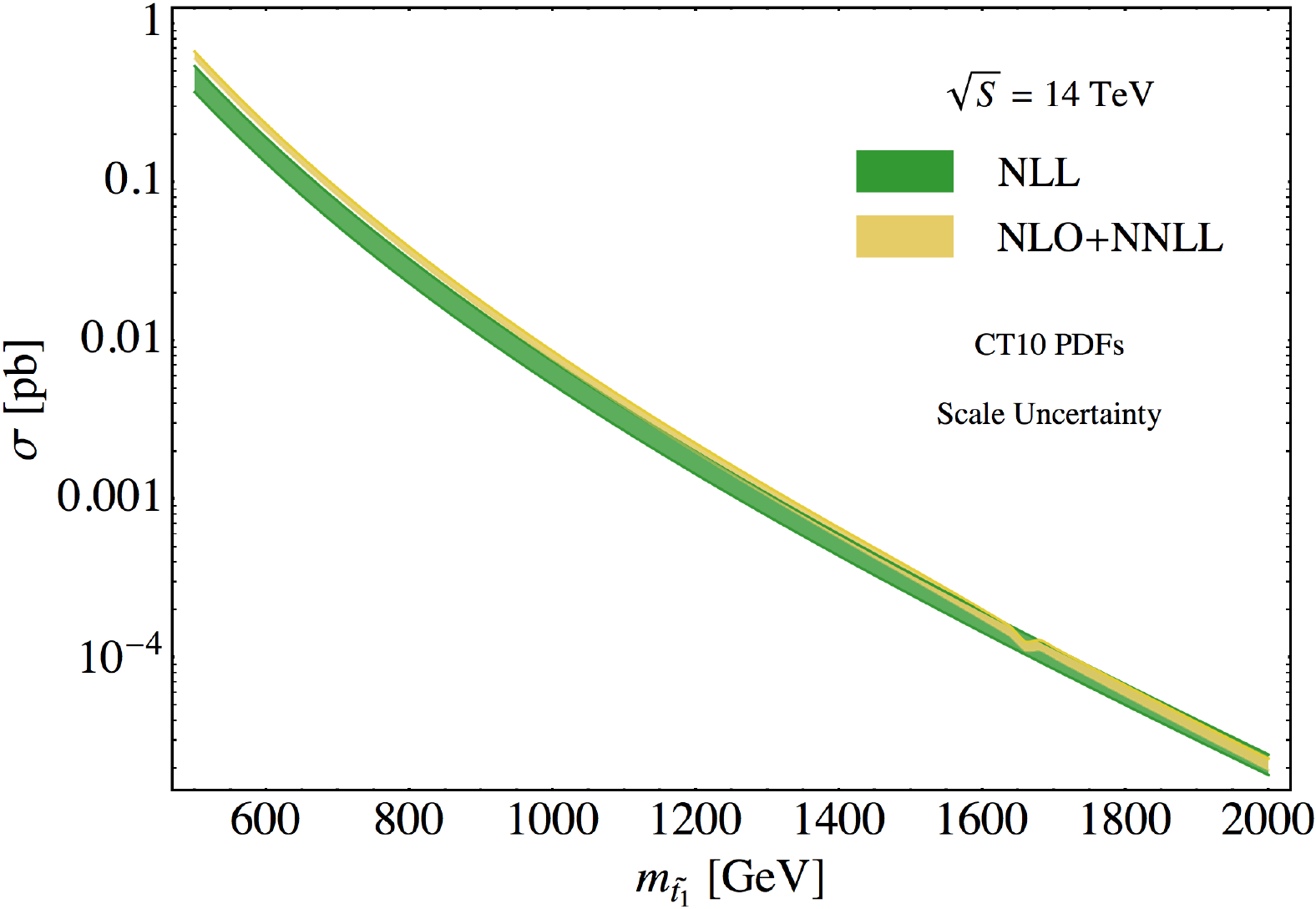} 
\end{tabular} 
\end{center}
\caption{\label{fig:massscanNLLvsNNLL} 
Comparison between NLL and NLO+NNLL predictions for the stop pair production cross section as a function of the top-squark mass. The first row refers to $\sqrt{S}=8$\,TeV, while the second row refers to $\sqrt{S}=14$\,TeV. All of the SUSY parameters other than $m_{\tilde{t}_1}$ are fixed at the values of the benchmark point {\tt 40.2.5} \cite{AbdusSalam:2011fc}. The plots are obtained by employing CT10NNLO PDFs.}
\end{figure}

This situation should be compared to the relation between approximate NNLO and NNLL 
+NLO predictions. While the total stop production cross section at NLO+NNLL accuracy is slightly smaller that the approximate NNLO cross section for all cases considered in the tables, the two predictions are well within the respective perturbative uncertainties, which are indicated by the first error next to the central values reported in the tables. The second error in the tables accounts for the PDF and $\alpha_s$ uncertainty. Both approximate NNLO and NLO+NNLL cross sections agree within perturbative uncertainties with the NLO calculations. The relative size of the approximate NNLO and NLO+NNLL corrections in the mass range $m_{\st} \in [500,2000]$\,GeV is 
shown in Figure~\ref{fig:NLOvcNNLL8TeVCT10}. As already observed in the tables, the approximate NNLO cross section is slightly larger than the NLO+NNLL one except in the case of very large $m_{\st}$ masses ($m_{\st} \gtrsim 1200$\,GeV for $\sqrt{S} = 8$\,TeV and $m_{\st} \gtrsim 1800$\,GeV for $\sqrt{S} = 14$\,TeV). The scale uncertainty of the predictions at NLO+NNLL accuracy is very similar to the scale uncertainty found in approximate NNLO calculations (around or smaller than 10 \%) in all cases analyzed in the tables. Both uncertainties are smaller than the corresponding NLO scale uncertainties and 
considerably
smaller than the corresponding PDF and $\alpha_s$ uncertainties. A comparison of the NLO and NLO+NNLL perturbative uncertainties in the stop mass range $m_{\st} \in [500,2000]$\,GeV is shown in Figure~\ref{fig:massscan8TCT10}, while the NLL and NLO+NNLL predictions for the total cross section in the same mass range are compared in Figure~\ref{fig:massscanNLLvsNNLL}. From the figure one can see that for low and moderate values of $m_{\st}$ the NLO+NNLL cross section is larger than the one obtained from calculations at NLL accuracy. The effect is particularly evident at $\sqrt{S} = 14$\,TeV. Figures~\ref{fig:massscan8TCT10} and \ref{fig:massscanNLLvsNNLL} show that the residual perturbative uncertainty  of NLO+NNLL calculations is smaller than the perturbative uncertainty affecting NLL and NLO calculation throughout the considered mass range. The tables and figures shown in this section indicate that the matched NLO+NNLL calculations improve the stability of the predictions for the stop pair production cross section.

These considerations serve as {\em a posteriori\/} self-consistency check of our calculational framework and indicate that the approximate NNLO and the NLO+NNLL predictions, which are in good agreement with each other, are robust. Of course, a full calculation of the NNLO corrections to the stop-pair production process would be the only way of assessing with certainty to which extent approximate NNLO calculations reproduce the exact NNLO results. Furthermore, NNLO calculations in fixed-order perturbation theory could be easily matched to the NNLL resummed cross section discussed in this work. Unfortunately to date the large number of mass scales involved makes a full evaluation of the NNLO corrections an extremely challenging task.

\subsection{Comparison with other results in the literature}

\begin{table}
\begin{center}
\begin{tabular}{|c|c|}
\hline  $\sqrt{S} = 14$\,TeV -- CTEQ6.6 PDFs  & $m_{\st} = 400$\,GeV \\ 
\hline
\hline $(\sigma+\Delta \sigma)_{\tNLO + \tNLL}$  [pb] from \cite{Beenakker:2010nq} & $21.9^{+2.0}_{-1.9} \times 10^{-2}$ \\ 
\hline $(\sigma+\Delta \sigma)_{\mbox{ {\tiny approx NNLO}}}$ [pb] from \cite{Broggio:2013uba} & $22.2^{+1.3}_{-1.0} \times 10^{-2}$ \\ 
\hline $(\sigma+\Delta \sigma)_{\tNLO + \tNNLL}$ [pb] this work & $21.3^{+1.0}_{-1.3} \times 10^{-2}$ \\ 
\hline 
\end{tabular} 
\end{center}
\caption{\label{tab:kulesza} 
Comparison between the NLO+NLL cross section of \cite{Beenakker:2010nq},  the approximate NNLO calculation of \cite{Broggio:2013uba}, and the NLO+NNLL result of the present work. The table refers to the LHC with $\sqrt{S} = 14$\,TeV and to $m_{\st} = 400$\,GeV, the remaining input parameters are set at the values characterizing the  SPS1a' benchmark point in \cite{AguilarSaavedra:2005pw}. The PDFs employed are the CTEQ6.6 set. We report only the perturbative uncertainty.}
\end{table}

\begin{table}[t]
\begin{center}
\begin{tabular}{|c||c|c|c|c|}
\hline
\multicolumn{5}{|c|}{LHC $7$\,TeV}  \\
\hline
\hline $m_{\st}$ [GeV]  & NLO  [pb] & NLL \cite{Falgari:2012hx} [pb] &  $\text{NNLO}_{\text{approx}}$\cite{Broggio:2013uba} [pb] & NLO+NNLL [pb]
 \\ 
 \hline
\hline $400$ & $0.211^{+0.028}_{-0.031}$ & $0.250^{+0.038}_{-0.030}$ & $0.226^{+0.011}_{-0.014}$ & $0.217^{+0.015}_{-0.014}$ \\ 
\hline $800$ & $1.09^{+0.16}_{-0.18}\times 10^{-3}$ & $1.34^{+0.20}_{-0.16}\times 10^{-3}$ & $1.22^{+0.04}_{-0.08}\times 10^{-3}$ & $1.20^{+0.12}_{-0.07}\times 10^{-3}$ \\ 
\hline $1000$ & $1.24^{+0.19}_{-0.21}\times 10^{-4}$ & $1.57^{+0.23}_{-0.18}\times 10^{-4}$ & $1.42^{+0.04}_{-0.10}\times 10^{-4}$ & $1.42^{+0.16}_{-0.09}\times 10^{-4}$ \\ 
\hline\hline
\multicolumn{5}{|c|}{LHC $8$\,TeV}  \\
\hline
\hline $m_{\st}$ [GeV]  & NLO  [pb] & NLL \cite{Falgari:2012hx} [pb] &  $\text{NNLO}_{\text{approx}}$\cite{Broggio:2013uba} [pb] & NLO+NNLL [pb] \\ 
\hline
\hline $400$ & $0.355^{+0.045}_{-0.051}$ & 0.416$^{+0.063}_{-0.050}$ & $0.378^{+0.020}_{-0.023}$ & $0.361^{+0.024}_{-0.021}$ \\ 
\hline $800$ & $2.48^{+0.33}_{-0.38}\times 10^{-3}$ & $3.00^{+0.43}_{-0.34}\times 10^{-3}$ & $2.73^{+0.08}_{-0.18}\times 10^{-3}$ & $2.67^{+0.26}_{-0.14}\times 10^{-3}$ \\ 
\hline $1000$ & $3.36^{+0.47}_{-0.54}\times 10^{-4}$ & $4.16^{+0.60}_{-0.47}\times 10^{-4}$ & $3.79^{+0.10}_{-0.25}\times 10^{-4}$ & $3.74^{+0.39}_{-0.22}\times 10^{-4}$ \\ 
\hline
\end{tabular} 
\end{center}
\caption{\label{tab:falgari} 
Comparison between the NLO+NLL results of \cite{Falgari:2012hx}, the approximate NNLO calculation of \cite{Broggio:2013uba}, and the NLO+NNLL results of the present work. The numbers refer to the benchmark point {\tt 40.2.4} \cite{AbdusSalam:2011fc}. In particular, we set $m_t = 172.5$\,GeV, $m_{\tilde{g}} = 1386$\,GeV, $m_{\tilde{q}} = m_{\tilde{t}_2} = 1358$\,GeV, and $\cos \alpha= 0.39$ 
as in \cite{Falgari:2012hx}. The numbers refer to the LHC operating at $\sqrt{S}=7$\,TeV (upper portion) and 8\,TeV (lower portion). The factorization scale is set equal to $m_{\st}$. For the approximate NNLO results and the NLO+NNLL results we used MSTW2008 NLO PDFs. The errors indicate only the perturbative uncertainty.}
\end{table}

We conclude our phenomenological analysis by comparing our NLO+NNLL predictions for the total cross section with the results obtained in \cite{Beenakker:2010nq} and \cite{Falgari:2012hx}, which have NLO+NLL accuracy and are obtained with calculational methods different from the ones employed here. We focus our attention on values of the stop mass which are close to or higher than the current lower bounds on this parameter in the MSSM.

In Table~\ref{tab:kulesza} we show the results obtained for the input parameters employed in \cite{Beenakker:2010nq}, which coincide with the SPS1a' benchmark point in \cite{AguilarSaavedra:2005pw}. In the table we consider a collider energy of $14$\,TeV and set $m_{\st} = 400$\,GeV. The PDF set employed is CTEQ6.6. We checked that, as expected, the NLO results in \cite{Beenakker:2010nq} coincide with the ones obtained with the {\tt Prospino} version we employ. Our central value for the NLO+NNLL cross section is  in very good agreement with the NLO+NLL value obtained in \cite{Beenakker:2010nq} and with the approximate NNLO prediction obtained in \cite{Broggio:2013uba}. The perturbative uncertainty of the NLO+NNLL result is essentially identical to the one affecting the  approximate NNLO calculation. Both are smaller than the NLO+NLL scale uncertainty.

Reference \cite{Falgari:2012hx} presents results obtained by resumming simultaneously production threshold logarithms and Coulomb singularities with NLL accuracy. Bound-state effects are also included in that calculation. Results for the top-squark pair production at the CMSSM benchmark point {\tt 40.2.40} \cite{AbdusSalam:2011fc} for $\sqrt{S} = 7$\,TeV and for several values of $m_{\st}$ are shown in the upper portion of Table~\ref{tab:falgari}. Coulomb resummation and bound state effects increase the cross section, but the largest effect in the NLL results of \cite{Falgari:2012hx} is due to soft resummation. A private version of the MSTW2008 NLO PDFs is employed in  \cite{Falgari:2012hx}, while in carrying out our calculations and comparisons we employ the public version of the same PDF sets. Since the NLO+NNLL results are very similar to the approximate NNLO calculations, the same observations made in \cite{Broggio:2013uba} apply also to the comparison of the NLO+NNLL results obtained here with the results of \cite{Falgari:2012hx}. In particular, one can see from the table that the NLO+NNLL predictions for the cross section are in good agreement with the NLL predictions once perturbative uncertainties are taken into account. The central values at NLO+NNLL accuracy are marginally smaller than in approximate NNLO calculations. The perturbative uncertainty is slightly larger than the one found at approximate NNLO, but smaller than the one quoted in \cite{Falgari:2012hx}. The lower portion of Table~\ref{tab:falgari} shows that the same observations apply to the case of the LHC at $\sqrt{S} = 8$\,TeV, for which the authors of \cite{Falgari:2012hx} provide predictions in an ancillary file included in the arXiv submission of their work.

Finally, we briefly comment on a few papers which were recently published. Ref.~\cite{Falgari:2012sq} analyzes the impact of finite-width effects on threshold corrections to squark and gluino production, finding them to be negligible for a moderate decay width, $\Gamma/m \leq 5\%$, which corresponds to the case of interest for present searches. This result confirms the validity of the analysis presented here, which neglects these effects. Refs.~\cite{Beenakker:2013mva,Beneke:2013opa} present the first results in threshold resummation (in the $\beta\to 0$ limit) for squark and gluino production at NNLL accuracy (the latter including Coulomb gluon effects as well). Since these papers focus on squark and gluino production and do not consider stop pair production, a direct comparison is not possible. It will be interesting to compare the different approaches when a comprehensive phenomenological analysis for stop pair production will be available.

\section{Conclusions}
\label{sec:Conc}

In this paper, we have completed the analysis of the soft-emission corrections to the production of top-squark pairs started in \cite{Broggio:2013uba}. In particular, we have considered the resummation of partonic threshold logarithms at NNLL order. Our method relies on the factorization of the partonic cross section in a trace of the product of two matrices, the hard and soft functions, in color space. This factorization is valid in the soft limit. The hard function accounts for virtual corrections, while the soft function accounts for the emission of soft gluons. In \cite{Broggio:2013uba}, it was shown that the use of the threshold limit of the partonic cross section allows one to obtain reliable predictions for hadronic observables in stop pair production, at least for the range of values of $m_{\st}$ considered in that work and in the present one. 
This happens because of the mechanism of dynamical threshold enhancement \cite{Becher:2007ty}, which essentially amounts to the fact that PDFs enhance the relative weight of the soft-emission region in the partonic phase-space integrals appearing in the calculation of hadron-initiated production process. Furthermore, in \cite{Broggio:2013uba} we presented the calculation of the hard and soft functions up to NLO and derived the anomalous dimensions of the hard and soft functions required in order to obtain approximate NNLO formulas for stop-pair production observables. These approximate formulas include all of the plus distributions appearing in the partonic cross section at NNLO, which capture the leading singular terms in the soft limit.

In the present work, we have solved the RGEs satisfied by the hard and soft functions in order to carry out the resummation of threshold logarithms directly in momentum space, with NNLL accuracy. The relevant anomalous dimensions are identical to the ones employed in the study of top-quark pair production considered in \cite{Ahrens:2010zv,Ahrens:2011mw}. We have carried out the analysis in two different kinematic schemes, PIM and 1PI, which in principle allow us to obtain different differential distributions, such as the stop-pair invariant-mass spectrum or the top-squark transverse-moment and rapidity distributions.

However, top squarks have not been discovered yet. Consequently, the most interesting observable in top-squark pair production is the total cross section, which must be evaluated as a function of the mass of the hypothetical top squark. Our technique allows us to obtain the total cross section by carrying out the resummation in either PIM or 1PI kinematics and subsequently integrating the double-differential distribution over the available phase-space. Furthermore, the difference between the predictions obtained in the two kinematic schemes provides a handle for how to estimate subleading corrections neglected in the soft limit. In fact, a different set of formally subleading corrections is neglected in the two different schemes. This scheme uncertainty is combined with the usual scale uncertainties in order to estimate the total perturbative uncertainty affecting our predictions.

The phenomenological predictions for the total cross section as a function of the top-squark mass have been obtained by matching the resummation formulas at NNLL order with the complete NLO calculation obtained from the code {\tt Prospino} \cite{Beenakker:1996ed}. The NLO+NNLL calculations  lead to values of the total cross section which are very close to the approximate NNLO calculations of the same observable, first presented in \cite{Broggio:2013uba}. The perturbative uncertainty affecting the NLO+NNLL calculations is essentially the same we found in approximate NNLO calculations; moreover, it is smaller than both the perturbative uncertainty affecting NLL calculations and the residual PDF and $\alpha_s$ uncertainty. We consider the good agreement between approximate NNLO and NLO+NNLL calculations an indication of the fact that our calculational framework is self consistent and robust, {\em a priori\/} equivalent to other schemes employed to carry out the resummation of soft gluon emission in this process. We note, however, that the resummation of higher-order Coulomb corrections, studied for example in \cite{Falgari:2012hx} at NLL accuracy, is not considered in the present work. 

Finally, we emphasize that the procedure described here and in \cite{Broggio:2013uba} can be adapted and applied to the study of other production processes involving colored supersymmetric particles, such as gluino pairs, sbottom pairs, and pairs of squarks of the first and second generation.

\subsection*{Acknowledgments}

The work of A.F.\ was supported in part by the PSC-CUNY Award No.\ 65214-00-43, by the PSC-CUNY Award No.\ 66590-00-44 and by the National Science Foundation Grant No.\ PHY-1068317. The research of M.N.\ is supported by the ERC Advanced Grant EFT4LHC of the European Research Council, the Cluster of Excellence {\em Precision Physics, Fundamental Interactions and Structure of Matter} (PRISMA -- EXC 1098) and grant NE 398/3-1 of the German Research Foundation (DFG), grant 05H12UME of the German Federal Ministry for Education and Research (BMBF), and the Rhineland-Palatinate Research Center {\em Elementary Forces and Mathematical Foundations}. The work of L.V. was supported by MIUR (Italy), under contracts 2006020509 004 and 2010YJ2NYW 006; by the Research Executive Agency (REA) of the European Union, through the Initial Training Network LHCPhenoNet under contract PITN-GA-2010-264564, and by the ERC grant 291377 ``LHCtheory: Theoretical predictions and analyses of LHC physics: advancing the precision frontier''. The research of L.L.Y. was supported in part by the National Natural Science Foundation of China under Grant No.\ 11345001. Computing resources were provided by the CTP cluster of New York City College of Technology (CUNY).

\bibliography{paperbib060312}
\bibliographystyle{JHEP-2}

\end{document}